# Active flow-driven DNA remodeling generates millimeter-scale mechanical oscillations

Maya Levanon[1], Noa S. Goldberg[1], Dvir Cohen[2], Eran Bouchbinder[1], Ram M. Adar[2], Alexandra M. Tayar[1*].

[1]Chemical and Biological Physics Department, Weizmann Institute of Science, Rehovot 7610001, Israel.

[2] Department of Physics, Technion, Israel Institute of Technology, Haifa 32000, Israel.

* e-mail: alexandra.tayar@weizmann.ac.il

**One-sentence summary:** A minimal active viscoelastic polymer composite converts local motor activity into synchronized, millimeter-scale oscillations through internal mechanical feedback, revealing a sharp nonequilibrium transition controlled by polymer network self-organization.

**Abstract**

**In living systems, DNA undergoes continuous and rhythmic mechanical remodeling through condensation, looping, and disentangling to regulate gene expression, segregate chromosomes, and guide morphogenesis. Here, we demonstrate a purely mechanical route to rhythmic DNA reorganization in a minimal active composite of microtubules, kinesin motors, and DNA. We embed a DNA polymer in an active turbulent microtubule-kinesin fluid, creating a self-morphing material. The active flows stretch and entangle the DNA, forming a self-organized viscoelastic network that resists active stresses and affects flow over large length scales. This mechanical feedback loop progressively amplifies velocity correlations and drives a nonequilibrium phase transition tuned by DNA contour length: from disordered flow to synchronized, millimeter-scale oscillations with vortices. We rationalize the phase transition with an active-gel model that predicts a growing length scale and an oscillatory instability emerging from the interplay between activity, orientational order, and self-generated viscoelasticity, rather than chemical signaling. The dependence of the oscillation frequency on system size and activity quantitatively agrees with experiment. Thus, flow-driven DNA remodeling provides a minimal physical route to autonomous, system-spanning oscillations in three dimensions and suggests design principles for programmable soft matter that coordinates, actuates, and reshapes itself.**

Introduction

Microtubule-molecular motor assemblies power essential physiological processes including the oscillatory chromosome movements during segregation (*1–3*), long-range intracellular transport in cytoplasmic streaming (*4, 5*), and locomotion through synchronous ciliary and flagellar beating (*6, 7*). These processes rely on active motor forces that give rise to coordinated dynamics spanning six orders of magnitude in time, from the rapid $10-50 Hz$ beat of respiratory cilia (*6, 8, 9*), to the slow $10^{-3} Hz$ waves that sculpt embryonic tissues (*7, 10*). Long-range coordination arises when motors are elastically linked to filaments via crosslinkers or an elastic matrix, enabling stress transmission across networks and synchronizing motion in space and time (*3, 11–13*). Yet, the fundamental mechanisms and spatial modes that underlie such global self-organized coordination remain poorly understood, largely due to the complexity of cellular systems (*14*).

Synthetic microtubule-kinesin motor systems have emerged as powerful reductionist platforms for shedding light on spatiotemporal modes of molecular motors-cytoskeleton filaments dynamics (*15–17*), revealing rich nonequilibrium behaviors including spontaneous flows (*18–21*), active turbulence (*22–24*), and confinement-induced coherent circulation (*25, 26*). Yet, without a mechanism for long-range elastic coupling, correlation lengths in these minimal systems remain finite, and no autonomous, phase-locked, system-scale oscillations were observed. Recent work has begun to probe active viscoelastic matter by introducing long-range elastic coupling into active systems, transforming local active constituent stresses into collective mechanical oscillations. Examples include reversible flow oscillations in confined bacterial-DNA suspensions (*27, 28*) and mechanically induced dense crowd oscillations under lateral confinement (*29*), both relying on external boundary conditions to stabilize the dynamics. More recently, crosslinked actin embedded in microtubule-based active matter has been shown to form thin elastic sheets that support propagating bending waves (*30*). The gap between synthetic active systems and autonomous biological coordination raises a fundamental question: What minimal components suffice to achieve autonomous biological-like mechanical coordination without such constraining geometry?

Here, we introduce a minimal three-dimensional microtubule (MT) based active composite in which elasticity is emergent. A DNA polymer embedded in a MT-kinesin active fluid acts as a flexible, viscoelastic scaffold once deformed by the active flow. As the flow stretches and entangles the DNA, a system-spanning network self-assembles, which in turn feeds back mechanically to couple distant regions and transform active turbulence into robust, period-selected, millimeter-scale oscillations. We uncover a sharp dynamical transition, set by activity and polymer length, that separates disordered active turbulence from coherent spatiotemporal order. The oscillatory state arose without chemical signaling, externally imposed orientational alignment rules, confining boundaries, or pre-built scaffolds. Rather, oscillations emerged from mechanical feedback between local motor stresses and a dynamically evolving, self-organized viscoelastic matrix. This establishes a minimal route to self-morphing soft materials that can sustain system-spanning oscillations in active matter.

**Results**

**Mechanical coupling in an active-passive viscoelastic composite drives a dynamical transition from turbulence to large-scale oscillations:** To test whether mechanical feedback can coordinate disordered active flows into system-scale oscillations, we built a composite biomaterial that couples a microtubule-kinesin active fluid to a passive, force-responsive DNA (Fig. 1A). The active fluid comprised MT filaments and kinesin molecular motor clusters (*31–33*). Poly(ethylene glycol) (PEG) induced depletion forces that bundled MTs, while kinesin clusters crosslinked and slid neighboring anti-aligned filaments, generating extensile, turbulent-like active flows with steady state velocities (*33, 34*) (Fig. S1). Into this fluid we introduced long, linear double stranded DNA, produced by concatenating $\lambda$-phage genomes to a tunable mean contour length of $L_{DNA} \sim 40 \mu m$ (Fig. S2, Methods). DNA behaves as an ideal flexible polymer (persistence length $\approx 50 nm$) with an elastic response in the range $1 - 65 pN$, matching kinesin-generated forces and enabling stress transmission over mesoscopic distances (*35–37*). Samples were loaded into a flow channel with height $H \approx 130 \mu m$, width $W \approx 2 mm$, length $L = 22 mm$, setting a length scale separation $H \ll W \ll L$ (Fig. S3). Initially, active MT flows mixed and stretched the DNA, promoting entanglement. Over time, the DNA spontaneously reorganized into a mesoscopic network that progressively coarsened (Fig. 1B). Fluorescent labeling of the DNA revealed a transition from a homogeneous solution to a well-defined network of coarse filaments (Fig. 1B, Cyan). Notably, the MTs did not assemble into mesoscale structures, and a passive DNA solution under same buffer and PEG depletion in the absence of MTs did not coarsen into a connected network (Fig. 1B, Gray, S4).

As the DNA network coarsened, the active fluid shifted from turbulent-like flow to sample-wide, temporally coherent positional oscillations (Fig. 1C,D). Passive tracer particle trajectories visualized network displacement and its evolution from local paths to 2mm-scale rotating domains (Video S1). Individual particle tracks showed oscillating displacements, with amplitudes growing to 0.25mm throughout the experiment's duration and confined within the 2mm rotating domain (Fig. 1D, S5). This confinement indicates periodic flow reversals that prevent tracers from completing full revolutions with the mm-scale vortex.

We quantified DNA-MT composite dynamics by applying particle image velocimetry (PIV) to MT images, as the velocity fields of DNA and MTs are indistinguishable within our experimental resolution (Methods, Fig. S6). Consistent with tracer trajectories, velocity heatmaps showed coherent rotating domains growing from local patches to 2 mm-scale vortices with increasing velocity magnitude, $v$ (Fig. 1E). These rotating domains periodically reversed direction on minute timescales, driving rhythmic displacement of the DNA network and yielding velocity oscillations. Complementary experiments spanning the full dynamical evolution demonstrated persistent oscillations in the coarse-grained in-plane velocity components, $v_x$ and $v_y$, with a characteristic period of $T = 25 \pm 9 min$ (Fig. 1F, Methods). Over time, the oscillation envelope grew, and the oscillations became more pronounced. Notably, DNA-free controls exhibited random flow with no preferred direction.

The velocity-velocity correlation length, $\xi_v$, obtained from an exponential fit of the spatial velocity autocorrelation function $C_{vv}(\vec{r},t) = \langle \vec{v}(\vec{r}_0,t) \cdot \vec{v}(\vec{r}_0+\vec{r},t)/v^2(\vec{r}_0,t)\rangle_{\vec{r}_0,\theta}$, increased sharply by an order of magnitude from ~$250\mu m$ to $\xi_{max} = 1.5mm$, approaching the channel width of $2mm$, before saturating (Fig. 1G, Methods). The increase in $\xi_v$ tracked DNA-network coarsening as manifested in a growing fluorescence-intensity coefficient of variance, While DNA-free controls remained constant at $\xi\_ max \approx 250\mu m$ (Fig. 1H, SI). These observations indicate that the active DNA-MT composite undergoes spontaneous self-organization, driven by the interplay between local motor-generated stresses and polymer network formation. Active flows progressively restructured the DNA into a mechanically connected scaffold, converting the composite from an active fluid to an elastically coupled, solid-like state that oscillated in unison. This dynamic feedback transformed the composite from a disordered, turbulent active fluid to an active viscoelastic gel transforming the dynamics from disordered turbulent flow to coherent, system-scale oscillations.

**Progressive DNA network coarsening extends correlation length and populates system-scale modes:** To understand how the DNA network achieves mechanical connectivity across the system scale, we examined its 3D structure by confocal microscopy, quantifying DNA distribution along the z-axis and relating it to the onset and scale of in-plane oscillations. Initially, DNA was homogenously distributed across the channel with no detectable structure (Fig. 2A). As the network formed and coarsened, DNA progressively condensed along the z-axis into a thin, quasi-2D sheet, as observed in cross-sectional images. We quantified the vertical contraction by fitting Gaussian functions to the z-axis fluorescence-intensity profiles, confirming that network thickness decreased over time (Fig. 2B, Methods).

Three temporal regimes emerged: (I) Early network formation with DNA evenly distributed, $0 < t < 60min$, during which oscillations already appear $t \approx 20min$ (Fig. 2C). (II) Continued coarsening with z-axis thinning, a concurrent increase in the velocity correlation length, $60 < t < 120min$ (Fig. 2D), and (III) a steady-state thickness $t > 120min$ in which $\xi_v$ saturates at $\xi_{max} \approx 1.5mm$ near the channel width, and the mean in-plane velocity magnitude $v$, recovers the DNA-free value $v_{active} \approx 3\mu m/s$ (Fig. 2D, S7, S8). The approach of $v$ to value $v_{active}$ indicates that, at long times, network reorganization permits high-magnitude, large-scale flows that are set by the intrinsic active dynamics of the MT-kinesin active fluid, with the DNA scaffold largely advected by, rather than strongly slowing, the MT-driven flow. Consistent with this picture, throughout this evolution the MT bundles show no detectable vertical reorganization and continue to span the channel height (Fig. 2A). MTs that are not incorporated into the dense composite network exhibit the same turbulent dynamics as in the DNA-free state.

Importantly, z-axis contraction is slow (~60$min$) compared with the oscillation period (~25$min$). Oscillations begin in regime I, before any vertical thinning, indicating that a mechanically connected 3D network, and not a fully condensed sheet, is sufficient. As thinning starts in regime II, $\xi_v$ rises sharply and long-range modes appear. The simultaneous increase of $\xi_v$ and mean velocity magnitude $v$ shows energy shifting into phase-coherent, long-wavelength modes that approach the channel scale. Together, these results show that mechanical connectivity in a 3D DNA network, rather than a fully condensed quasi-2D sheet, is sufficient to support system-spanning oscillations.

To characterize how motion is organized across spatial scales, we computed the isotropic velocity power spectrum of the MT velocity field $E(k)$, accumulated over 10$min$ windows (Fig. 2E, S9, Methods). Here, $E(k) \propto \frac{k}{2}\langle v(k)^2 \rangle$ so that $\int E(k)\,dk = \frac{1}{2}\langle |v|^2 \rangle$ and $v(k)$ is the spatial Fourier transform of the velocity magnitude. Before large-scale oscillations emerged at $t = 30min$, the spectrum plateaued at intermediate wavenumbers $k = 2.2 \cdot 10^{-3} \mu m^{-1}$ compatible with $\xi_v = 450 \mu m$, and comparable to the scale of the plateau of DNA-free MT active fluids (*22, 23*). As the DNA network coarsened at $T > 100min$, spectral power increased across all $k$ with a pronounced shift toward low-$k$, while high-$k$ powers persisted. This shows the emergence of system-scale vortices on top of sustained local activity, and a net build-up of kinetic energy. After normalizing spectra by their instantaneous minima, curves from all times collapse to a common scaling $E(k) \propto k^{-1.85}$ over a scale of $100\mu m - 2mm$, this contrasts with the $E(k) \propto k^{-1}$ behavior typical of DNA-free MT active fluids which is observed only over the narrower range of $30 - 200\mu m$ (*22, 23*) (Fig. S9). Thus, the distribution of energy across scales becomes self-similar, while the steeper slope indicates that energy accumulation preferentially amplifies large-scale modes without extinguishing small-scale fluctuations.

Next, we asked how kinetic energy is distributed across scales between the MT flow and the DNA network (Fig. 2F). To resolve between the MT-kinesin and DNA velocity fields, we performed complementary high-magnification, fast measurements targeting wavelengths $\lambda = k^{-1} \approx 10 - 600\mu m$, and directly compared mode-resolved power spectra of MT (solid line) and DNA (dashed line) velocities in the same field of view over 4-min windows. Interestingly, the MT spectra exceed the DNA spectra on intermediate scales $k^{-1} \approx 10 - 300\mu m$, while at longer wavelengths the two velocity fields are similar within our measurement sensitivity. This is consistent with permeation that occurs on small scales. As the DNA network forms and coarsens, the DNA power at these intermediate modes increases and, after a lag of ~10$min$, comparable to roughly half the oscillation period, catches up to and thereafter tracks the MT spectra during and after network formation. This scale-specific delay shows that motor-driven active stresses are first injected into the fluid and only later taken up by the assembling DNA scaffold, setting the timescale over which viscoelastic feedback between the active flow and the network is established.

**Polymer length drives a nonequilibrium phase transition from active turbulence to synchronized oscillations:** Having observed a polymer-mediated transition from active turbulence to coherent oscillations, we next asked whether the contour length of DNA determines the onset threshold for network formation and the emergence of long-range order. To control the DNA length distribution, we covalently ligated the cohesive ends of $\lambda$-DNA genomes using T4 DNA ligase in the presence of capping oligonucleotides that block the ends (Fig. 3A, Methods). By varying the DNA-to-cap molar ratio at a fixed DNA concentration, we produced length distributions of concatemers spanning ~1-10 monomers, corresponding to contour lengths ~$15 - 150 \mu m$, (Fig. 3B, S2). Each DNA solution was combined with the MT-kinesin active fluid at a fixed DNA concentration of $70 ng/\mu l$. Finally, the maximal velocity-velocity correlation length $\xi_{max}$ was measured to assess the degree of long-range order.

Varying the average DNA contour length $L_{DNA}$ revealed a sharp transition in the flow regime of the active composite. The maximal correlation length, $\xi_{max}$, exhibited a discontinuous, first-order-like, increase by an order of magnitude from ~$250 \mu m$ to ~$2000 \mu m$, as a function of $L_{DNA}$. This jump delineates two steady-state regimes separated by a metastable transient mesoscale state, each with distinct flow organization and characteristic DNA structure (Fig. 3C,D,S10A): (i) **Active turbulence regime:** ($L_{DNA} < 29 \mu m$) short DNA produced turbulent, short-range-correlated flows with homogeneously distributed DNA (Fig. S10B, Video S2,S3). (ii) **Metastable vortices ($L_{DNA} \approx 29 - 39 \mu m$)**: intermediate DNA lengths produced a transient array of counter-rotating vortices that flow unidirectionally, with a lifetime of $60 min$ that subsequently decayed back to active turbulence (Fig. 3E, yellow). The vortex dynamics was accompanied by the formation of large, discontinuous DNA strings (Video S4,S5). (iii) **Oscillatory regime:** ($L_{DNA} > 39 \mu m$): long DNA sustained persistent, synchronized velocity oscillations with millimeter-scale vortices, and a continuously coarsening DNA network (Video S6,S7). Unlike the transient state, the flow forms a stable lattice in which individual vortices roll along the channel's long axis and are continuously replaced by counter-rotating vortices of the same size, with occasional direction switches (Fig. 3E, blue, S11). This rolling dynamic produces coherent oscillatory patterns with phase speed comparable to the steady state mean velocity. Space-time plots of $v_x$ and $v_y$ revealed propagating phase bands of a phase offset compatible with vortex rolling along the channel long axis at a system saturated velocity comparable to $v_{active}$. Tracer particle trajectories formed closed ellipses with negligible net drift, confirming oscillatory motion is a wave without bulk transport (Fig. S5).

The sharp transition between the active turbulence and oscillations regimes is controlled by the DNA length, whether plotted against mean or maximal contour lengths (Fig. S12), implying that the transition is robust with respect to the underlying length distribution. Moreover, varying the short DNA concentration did not trigger a transition, and the flows remained active-turbulent, reinforcing the central role of the polymer length (Fig S13). Qualitatively, the data indicate a geometric length-matching condition: when DNA lengths (maximal $L_{DNA} \approx 100 \mu m$ compatible with mean $L_{DNA} \approx 40 \mu m$), approach the initial activity-set correlation length $\xi_v(t = 0) = 100 - 200 \mu m$ (33) (Fig. 1G, S1, S12), the stretched polymers bridge inter-vortex gaps, entangle, and assemble a system-spanning DNA network (Fig. 1B). As the polymer network dynamically forms, it acquires viscoelastic properties on progressively increasing length scales, enabling the formation of coherent spatiotemporal structures. Specifically, it features a viscoelastic relaxation time $\tau_{VE}$, which is expected to increase with the polymer length. Next, we develop a theoretical model inspired by this picture, which yields quantitative, testable predictions.

**A minimal model of a growing correlation length and the onset of oscillations:** We model the active MT-DNA composite as an active viscoelastic nematic gel (*27, 38*) (SI). We focus on the two-dimensional (2D) in-plane dynamics of the three-dimensional (3D) system in view of the system's geometry that satisfies $H \ll W, L$. This effective description leads to a hydrodynamic frictional force density $\Gamma \vec{v}$, were $\Gamma$ is the associated friction-like coefficient and $\vec{v}$ is the in-plane velocity. This force is balanced by variations in the in-plane stresses, including both passive viscoelastic and active nematic contributions. The elastic stress evolves according to the polymer network viscoelasticity, characterized by the relaxation time $\tau_{VE}$ and an elastic modulus. The nematic director, describing the local MT orientation, evolves similarly to the minimal description of previous work that includes a coupling between the director and the network strain, with a characteristic strain-alignment timescale $\tau_R$ (*27, 30*) (SI).

Our analysis first considers the dependence of the 2D in-plane velocity correlation length $\xi_v$ on the long-time viscosity and timescales ratio $\tau_{VE}/\tau_R$. Qualitatively, a large $\tau_{VE}/\tau_R$ indicates that elastic stress reorients the director of the MTs before it relaxes. In our system, both the viscosity and $\tau_{VE}/\tau_R$ are expected to increase with the polymer length. By analyzing the linear instability that drives the active flow (SI, Eq. 13), we find that $\xi_v$ increases monotonically with $\tau_{VE}/\tau_R$ and the viscosity. The first modifies the activity-driven alignment of MTs and the latter generally calms the dynamics. In a finite system, $\xi_v$ cannot increase indefinitely, hence we expect it to saturate at the smallest in-plane geometrical dimension, assuming the latter is smaller than the hydrodynamic screening length. Since in our system $W \ll L$, we expect $\xi_v \sim W$. Next, we show that the same model gives rise to yet another instability at sufficiently large $\tau_{VE}$ values leading to large-scale oscillations governed by the strain-alignment timescale $\tau_R$ and the magnitude of the motor-driven active stress $\alpha$ (SI, section 3.3). These oscillations feature a period $T$ that scales as $T \sim W\sqrt{\frac{\Gamma \tau_R}{\alpha}}$ (SI. Eq. 17). This scaling relation is in line with the expectation that $T$ should increase with both $\tau_R$ and the active timescale $\tau_0 \sim \Gamma \xi_v^2/\alpha$, resulting in $T \sim \sqrt{\tau_R \tau_0}$. The latter recovers the expression above once we set $\xi_v \sim W$. The oscillations period scaling $T \sim W/\sqrt{\alpha}$ is next tested experimentally.

**Oscillation period is set by geometry-limited correlation length and kinesin-driven flow velocity:** To test the theoretical prediction for the oscillation period $T$, we perturbed geometry and activity while fixing the DNA contour length at $L_{DNA} = 40\mu m$, well within the oscillatory regime where network formation is robust. We first varied the channel width $W$ at fixed motor concentration $C_{motor} = 267nM$. As $W$ increased, $\xi_{max}$ grew linearly without saturation over a tenfold range, $0.5 - 5\ mm$, indicating that width, rather than hydrodynamic screening, sets the largest accessible correlation scale (Fig. S14). In parallel, the MT-DNA network composite average velocity of magnitude $v$ increased with $W$, and approached a maximum at $W > 3mm$, suggesting that wall drag limits flow in narrow channels and becomes negligible in wider ones. To verify that intrinsic active stress is geometry-independent, we measured microtubule velocity in DNA-free samples, $v_{active}$, which remained constant across widths (Fig. S14B). Consistent with the theoretical scaling $T \propto W/\sqrt{\alpha}$, we find that the oscillation period increased linearly with $W$ at fixed activity (Fig. 4A). Space-time plots show diagonal bands of growing spacing with $W$, reflecting longer periods (Fig. 4B). Thus, geometry sets the upper bound on the accessible mode, and therefore the maximum period.

Second, to test the predicted scaling $T \propto 1/\sqrt{\alpha}$, we varied motor concentration, $C_{motor}$, at fixed channel width $W = 2mm$, thereby tuning activity while holding the geometric upper bound on $\xi_{max}$ constant (Fig. S15). In active turbulence, the intrinsic flow speed scales as $v_{active} \propto \sqrt{\alpha}$ (34), which allows us to rewrite the theoretical prediction as an experimentally testable relation between the oscillation period and the measured saturated velocity: $T \propto 1/v_{active}$. This approach provided approximately 2 orders of magnitude in dynamic range, with $v$ increasing from $0.1 \mu m/s$ to $8 \mu m/s$ (Fig. S15C). At the lowest activity tested $C_{motor} = 134 nM$, no visible DNA network formed, the correlation length remained short, and only transient, localized oscillations were observed (Fig. S15). Above this threshold, a DNA network emerged and progressively coarsened with increasing motor concentration, and $\xi_{max}$ remained close to the confinement limit and exhibited no measurable dependence on $v_{active}$ (Fig. S15B). Plotting $T$ as a function of $v_{active}$ reveals the expected inverse relationship $T \propto 1/v_{active}$, confirming the theoretical scaling (Fig. 4C). Thus, when $\Gamma$ and $\tau_R$ are approximately fixed, the data are consistent with an advection-time interpretation according to which the period is the turnover time of a vortex of size $T \propto \xi_{max}/v_{active}$, matching the experimentally measured period of $T \sim 20 min$ for $\xi_{max} \approx 2mm$, $v \approx 2 \mu m/s$ (Fig. S15D).

**Discussion**

Active matter can build its own ruler of space and time. In our composite, motor-driven MT flows stretch and entangle DNA into a viscoelastic scaffold that then feeds back to organize the flow. This self-organization is controlled by a geometric length-matching condition, where the typical DNA contours become comparable with the activity-set correlation length. The emergent viscoelasticity drastically alters the dynamics. The DNA network stores active stress and releases it with a finite delay, phase-locking distant regions into a common rhythm without biochemical cues or templating boundaries. In addition to providing delayed mechanical feedback, the network also mediates orientational coupling between local strain and the active MT field. The strength of this mechanism is quantified by the timescales ratio $\tau_{VE}/\tau_R$, comparing the duration of elastic memory to the timescale for strain-induced alignment.

The above picture is reinforced by two main spatiotemporal observations. First, timescale separation: oscillations appear while the DNA remains three-dimensional and only weakly condensed, whereas network thickness and correlation length evolve over a much longer timescale. Thus, coherence does not require a pre-formed sheet, it requires mechanical connectivity and a sufficiently large viscoelastic relaxation time so that elastic memory outlasts MTs reorientation and can drive long-range alignment, consistently with the theoretical model presented. Second, energy is organized across scales such that motors initially inject power at short wavelengths, after a finite lag, the DNA network catches up, and kinetic energy progressively populates longer wavelengths while small-scale motion persists. This redistribution of energy, which is consistent with the theoretical prediction of a growing length scale, yields a material that times itself and coordinates over millimeters, while staying vigorously active locally.

The system's failures to reach coherence are equally instructive. With sufficiently long DNA, a network forms, $\xi_v$ grows to the geometric limit, and a global oscillatory mode emerged with an observed mean velocity increase. With short polymers, the network cannot form, and flows remain actively turbulent and short-range correlated. Between these regimes we observe metastable, mesoscale rotations that decay once connectivity fails to span the system, a behavior expected when the length-matching condition is near, but not fully, met. Crucially, activity rewrites polymer physics as motor-driven stretching and alignment lower the effective entanglement threshold, creating nonequilibrium connectivity that would not arise at equilibrium.

Finally, the mechanism is general. Any semiflexible or flexible polymer that can be stretched within the molecular motor force range should, under drive, entangle and form a network, closing the same negative-feedback loop with an elastic delay and reorientation to produce synchronized oscillations. The design rule is simple: match polymer contour length to the activity-set correlation length and ensure $\tau_{VE}/\tau_R$ is large enough for viscoelastic memory to beat reorientation. Geometry then fixes $\xi_v$ (and therefore maximum $T$), while activity sets the pace. This perspective opens directions to designing programmable, autonomous soft materials, built from active fluids and drive-responsive polymers.

**Authors contribution**

M.L., N.S.G., and A.M.T. designed and performed the experiments and analyzed the data. D.C., E.B., and R.M.A. developed the theory. A.M.T., M.L., D.C., E.B., and R.M.A. wrote the manuscript. All authors edited and approved the final version of the manuscript.

**Acknowledgments**

A.M.T. acknowledges support from the Israel Science Foundation (ISF, Grant No. 1404/23) and the Alon Fellowship. AMT acknowledges that this research was made possible in part by the historic generosity of the Harold Perlman Family. A.M.T. also thanks Sattvic Ray for fruitful discussions that contributed to the formation of this work. R.M.A. acknowledges support from the Israel Science Foundation (ISF, Grant No. 444/25). E.B. acknowledges support from the Ben May Center for Chemical Theory and Computation and from the Harold Perlman Family.

**References**

1. S. W. Grill, J. Howard, E. Schäffer, E. H. K. Stelzer, A. A. Hyman, The distribution of active force generators controls mitotic spindle position. *Science (1979)* **301**, 518–521 (2003).

2. S. W. Grill, P. Gönczy, E. H. K. Stelzer, A. A. Hyman, Polarity controls forces governing asymmetric spindle positioning in the caenorhabditis elegans embryo. *Nature* **409**, 630–633 (2001).

3. H. Y. Wu, G. Kabacaoğlu, E. Nazockdast, H. C. Chang, M. J. Shelley, D. J. Needleman, Laser ablation and fluid flows reveal the mechanism behind spindle and centrosome positioning. *Nature Physics* **20**, 157–168 (2024).

4. S. Ganguly, L. S. Williams, I. M. Palacios, R. E. Goldstein, Cytoplasmic streaming in Drosophila oocytes varies with kinesin activity and correlates with the microtubule cytoskeleton architecture. *Proceedings of the Natural Academy of Science USA* **109**, 15109–15114 (2012).

5. R. E. Goldstein, J. W. van de Meent, A physical perspective on cytoplasmic streaming. *Interface Focus* **5** (2015).

6. F. Jülicher, J. Prost, Spontaneous Oscillations of Collective Molecular Motors. *Physics Review Letters* **78**, 4510 (1997).

7. S. Camalet, F. Jülicher, Generic aspects of axonemal beating. *New Journal of Physics* **2**, 324 (2000).

8. K. Kruse, F. Jülicher, Oscillations in cell biology. *Current Opinion in Cell Biology* **17**, 20–26 (2005).

9. F. Jülicher, J. Prost, Cooperative Molecular Motors. *Physics Review Letters* **75**, 2618 (1995).

10. G. Singer, T. Araki, C. J. Weijer, Oscillatory cAMP cell-cell signaling persists during multicellular Dictyostelium development. *Communications Biology* **2**, 1–12 (2019).

11. T. Ishikawa, T. Torisawa, H. Wada, A. Kimura, Swirling Instability Mediated by Elastic and Hydrodynamic Couplings in Cytoplasmic Streaming. *Physics Review X Life* **3**, (2025).

12. G. De Canio, E. Lauga, R. E. Goldstein, Spontaneous oscillations of elastic filaments induced by molecular motors. *Journal of the Royal Society Interface* **14** (2017).

13. J. C. Nawroth, A. M. Van Der Does, A. Ryan, E. Kanso, Multiscale mechanics of mucociliary clearance in the lung. *Philosophical Transactions of the Royal Society B: Biological Sciences* **375**, (2019).

14. F. Jülicher, Mechanical oscillations at the cellular scale. *Comptes Rendus de l'Académie des Sciences - Series IV - Physics-Astrophysics* **2**, 849–860 (2001).

15. M. C. Marchetti, J. F. Joanny, S. Ramaswamy, T. B. Liverpool, J. Prost, M. Rao, R. A. Simha, Hydrodynamics of soft active matter. *Reviews of Modern Physics* **85**, 1143 (2013).

16. R. Zhang, N. Kumar, J. L. Ross, M. L. Gardel, J. J. De Pablo, Interplay of structure, elasticity, and dynamics in actin-based nematic materials. *Proceedings of the Natural Academy Science USA* **115**, E124–E133 (2017).

17. T. D. Ross, H. J. Lee, Z. Qu, R. A. Banks, R. Phillips, M. Thomson, Controlling organization and forces in active matter through optically defined boundaries. *Nature* **572**, 224–229 (2019).

18. T. Sanchez, D. Welch, D. Nicastro, Z. Dogic, Cilia-like beating of active microtubule bundles. *Science (1979)* **333**, 456–459 (2011).

19. P. Chandrakar, M. Varghese, S. A. Aghvami, A. Baskaran, Z. Dogic, G. Duclos, Confinement Controls the Bend Instability of Three-Dimensional Active Liquid Crystals. *Physics Review Letters* **125**, (2020).

20. G. A. Vliegenthart, A. Ravichandran, M. Ripoll, T. Auth, G. Gompper, Filamentous active matter: Band formation, bending, buckling, and defects. *Science Advances* **6** (2020).

21. R. Aditi Simha, S. Ramaswamy, Hydrodynamic Fluctuations and Instabilities in Ordered Suspensions of Self-Propelled Particles. *Physics Review Letters* **89**, (2002).

22. B. Martínez-Prat, R. Alert, F. Meng, J. Ignés-Mullol, J. F. Joanny, J. Casademunt, R. Golestanian, F. Sagués, Scaling Regimes of Active Turbulence with External Dissipation. *Physics Review X* **11**, (2021).

23. R. Alert, J. F. Joanny, J. Casademunt, Universal scaling of active nematic turbulence. *Nature Physics* **16**, 682–688 (2020).

24. A. J. Tan, E. Roberts, S. A. Smith, U. A. Olvera, J. Arteaga, S. Fortini, K. A. Mitchell, L. S. Hirst, Topological chaos in active nematics. *Nature Physics* **15**, 1033–1039 (2019).

25. K. T. Wu, J. B. Hishamunda, D. T. N. Chen, S. J. DeCamp, Y. W. Chang, A. Fernández-Nieves, S. Fraden, Z. Dogic, Transition from turbulent to coherent flows in confined three-dimensional active fluids. *Science (1979)* **355** (2017).

26. J. Hardoüin, R. Hughes, A. Doostmohammadi, J. Laurent, T. Lopez-Leon, J. M. Yeomans, J. Ignés-Mullol, F. Sagués, Reconfigurable flows and defect landscape of confined active nematics. *Communications Physics* **2**, 1–9 (2019).

27. S. Liu, S. Shankar, M. C. Marchetti, Y. Wu, Viscoelastic control of spatiotemporal order in bacterial active matter. *Nature* **590**, 80–84 (2021).

28. Y. Qiang, C. Luo, D. Zwicker, Nonlocal Elasticity Yields Equilibrium Patterns in Phase Separating Systems. *Physics Review X* **14**, 021009 (2024).

29. F. Gu, B. Guiselin, N. Bain, I. Zuriguel, D. Bartolo, Emergence of collective oscillations in massive human crowds. *Nature* **638**, 112–119 (2025).

30. J. Berezney, S. Ray, I. Kolvin, F. Brauns, S. Chen, M. Bowick, S. Fraden, V. Vitelli, Z. Dogic, Active assembly and non-reciprocal dynamics of elastic membranes. arXiv: 2408.14699 (2024).

31. T. Sanchez, D. T. N. Chen, S. J. Decamp, M. Heymann, Z. Dogic, Spontaneous motion in hierarchically assembled active matter. *Nature* **491**, 431–434 (2012).

32. S. J. DeCamp, G. S. Redner, A. Baskaran, M. F. Hagan, Z. Dogic, Orientational order of motile defects in active nematics. *Nature Materials* **14**, 1110–1115 (2015).

33. G. Henkin, S. J. DeCamp, D. T. N. Chen, T. Sanchez, Z. Dogic, Tunable dynamics of microtubule-based active isotropic gels. *Philosophical Transactions of the Royal Society A: Mathematical, Physical and Engineering Sciences* **372** (2014).

34. L. Giomi, Geometry and topology of Turbulence in active nematics. *Physics Review X* **5**, (2015).

35. A. S. Backer, A. S. Biebricher, G. A. King, G. J. L. Wuite, I. Heller, E. J. G. Peterman, Single-molecule polarization microscopy of DNA intercalators sheds light on the structure of S-DNA. *Science Advances* **5** (2019).

36. A. M. Tayar, M. F. Hagan, Z. Dogic, Active liquid crystals powered by force-sensing DNA-motor clusters. *Proceedings of the National Academy of Sciences* **118** (2021).

37. K. Hall, D. G. Cole, Y. Yeh, J. M. Scholey, R. J. Baskin, Force-velocity relationships in kinesin-driven motility. *Nature* **364**, 457–459 (1993).

38. S. P. Thampi, R. Golestanian, J. M. Yeomans, Velocity Correlations in an Active Nematic. *Physics Review Letters* **111**, (2013).

39. T. Kozhukhov, B. Loewe, T. N. Shendruk, Mitigating Density Fluctuations in Particle-based Active Nematic Simulations. *Communications Physics* **7** (2024).

**Figures:**

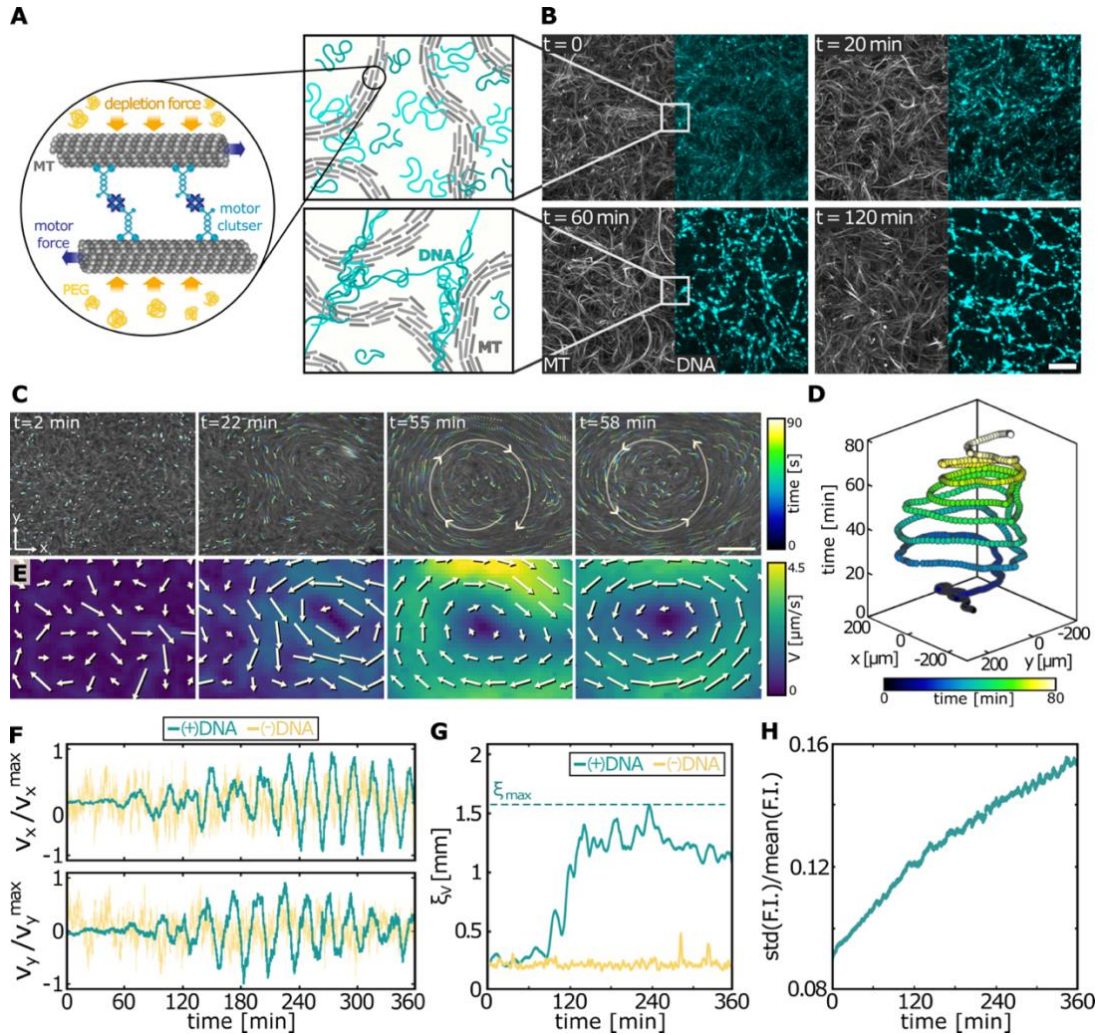

**Figure 1: Large-scale mechanical oscillations emerge from interactions of passive DNA polymers with microtubule-based active matter**. (A) Schematic of a microtubule-kinesin active system (gray) embedded in a DNA polymer (cyan). Motor-driven microtubule activity generates flows that extend and entangle the DNA. Inset: Kinesin clusters exert shear by sliding PEG-induced microtubule bundles. (B) Time-lapse fluorescent images of labeled DNA (cyan) embedded in a microtubule-based active gel (gray), during network formation and coarsening. Images are maximum-intensity z-projections from a confocal stack. Scale bar, $200\mu m$. (C) Time lapse of tracer particles trajectories and microtubules (gray) over $90\ sec$ (time colored), with overlaid vortex and orientation vectors as a guide to the eye. Scale bar, $400\mu m$. (D) Individual particle trajectory over 80 min. (E) Velocity magnitude heatmaps and corresponding vector fields for the data presented in C. (F) Velocity as a function of time at a fixed position in the MT-based active material, with and without DNA, averaged over a $130\mu m \times 130\mu m$ region. Velocities are normalized by their maximal value. (G) Velocity-velocity correlation length over time for samples with and without DNA. For (F,G) Acquisition starts earlier than in (C,E), capturing the full dynamics. (H) Normalized coefficient of variation of DNA fluorescence intensity, corresponding to the data in B and F-G, computed as the standard deviation divided by the mean intensity. All experiments were conducted at a DNA concentration of $70ng/\mu l$ and mean DNA contour length of $\sim 40\mu m$. Note that, due to wavelength overlap, MTs are imaged either with DNA or with beads in separate experiments; accordingly, the data in C-E were acquired in a complementary experiment distinct from that in F-H.

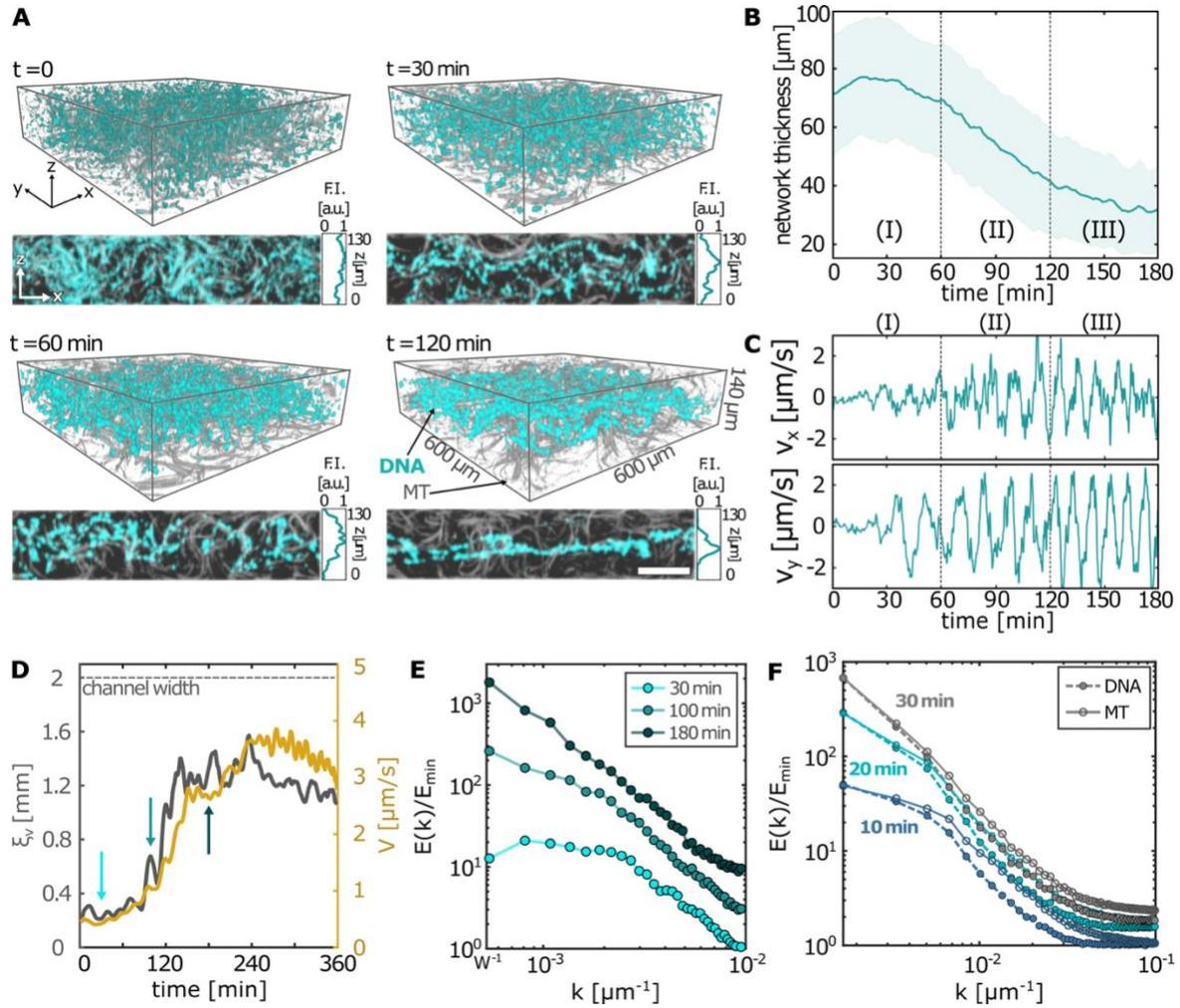

**Figure 2: Polymer network organization drives large-scale motion.** (A) 3D confocal time lapse of the MT active gel (gray) embedded in DNA polymers (cyan), showing DNA reorganization to form a network and contract into a 2D sheet. Volume and side views are shown. Scale bar, $100 \mu m$. (B) DNA network thickness as a function of time, obtained by fitting fluorescence intensity profiles with Gaussian functions (Methods). The shaded region indicates the standard error of the fitted widths across 65×65 μm² tiles. Dashed lines indicate the regimes indicated in the text. (C) Velocity over time along the X-Y plane. Velocity fields were obtained from maximum-intensity z-projections of DNA confocal images. (D) Field average of velocity-velocity correlation length (gray) and mean composite velocity magnitude (yellow) over time. Arrows mark times corresponding to the curves in (E). (E) Normalized kinetic energy spectrum of MT as a function of wave number $k$ at three time points. The geometry-limited cutoff is marked as $W^{-1}$. (F) Normalized kinetic energy spectrum of MT and DNA as a function of wave number $k$ at three early time points derived from the complementary high-magnification view. Note that panel F shows data from a separate, higher-magnification experiment than panel E; accordingly, the absolute velocity magnitudes differ between the two.

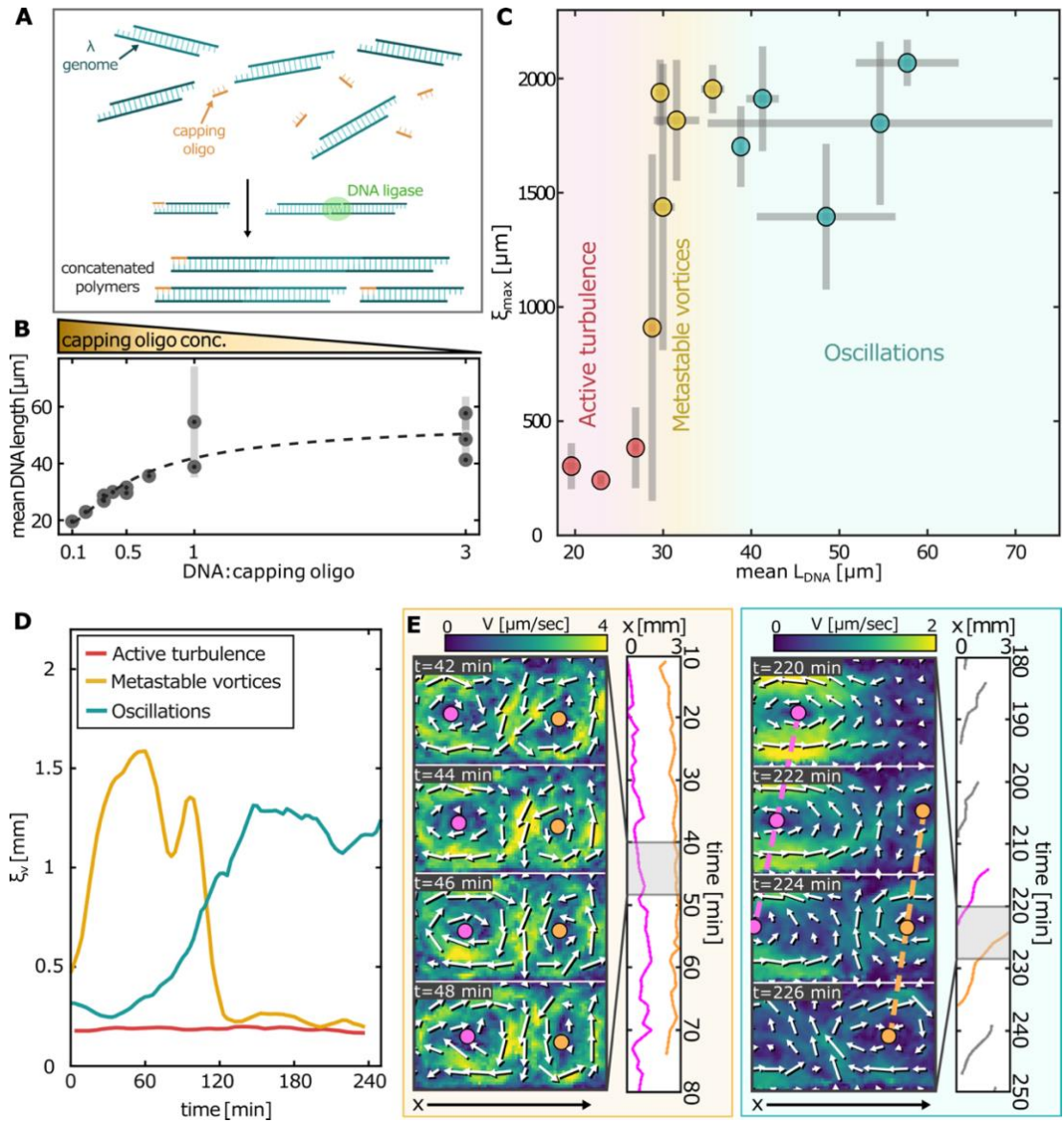

**Figure 3: Polymer length drives a sharp transition in correlation length.** (A) Schematic of λ-phage DNA concatenation by ligase, limited by capping oligonucleotides. (B) Weighted-average polymer length after concatenation as a function of the λ-genome to capping-oligonucleotide ratio. The error bars show the standard deviation over three replicate gel lanes. Polymer length distributions shown in Fig. S2. (C) Maximal velocity-velocity correlation length as a function of polymer length. Colors denote three flow regimes: red: Local active turbulence, yellow: Metastable vortices, blue: Oscillation. Horizontal error bars show the standard deviation of mean polymer length over 3 replicates, corresponding to the error bars in 3B. Vertical error bars indicate standard deviation of maximal velocity-velocity correlation length over 3-5 repeats. (D) Examples of velocity-velocity correlation length as a function of time for the flows described in (C). (E) Time lapse of velocity magnitude heat maps in the metastable vortices (yellow) and oscillatory (blue) regimes, overlaid with arrow indicating the velocity vector fields. Vortex centers are indicated on image. Right insets: vortex center position as a function of time at different time points. Colored tracks correspond to vortex trajectories shown in the heatmaps. A complementary plot of vortex velocity over time is presented in Fig. S10.

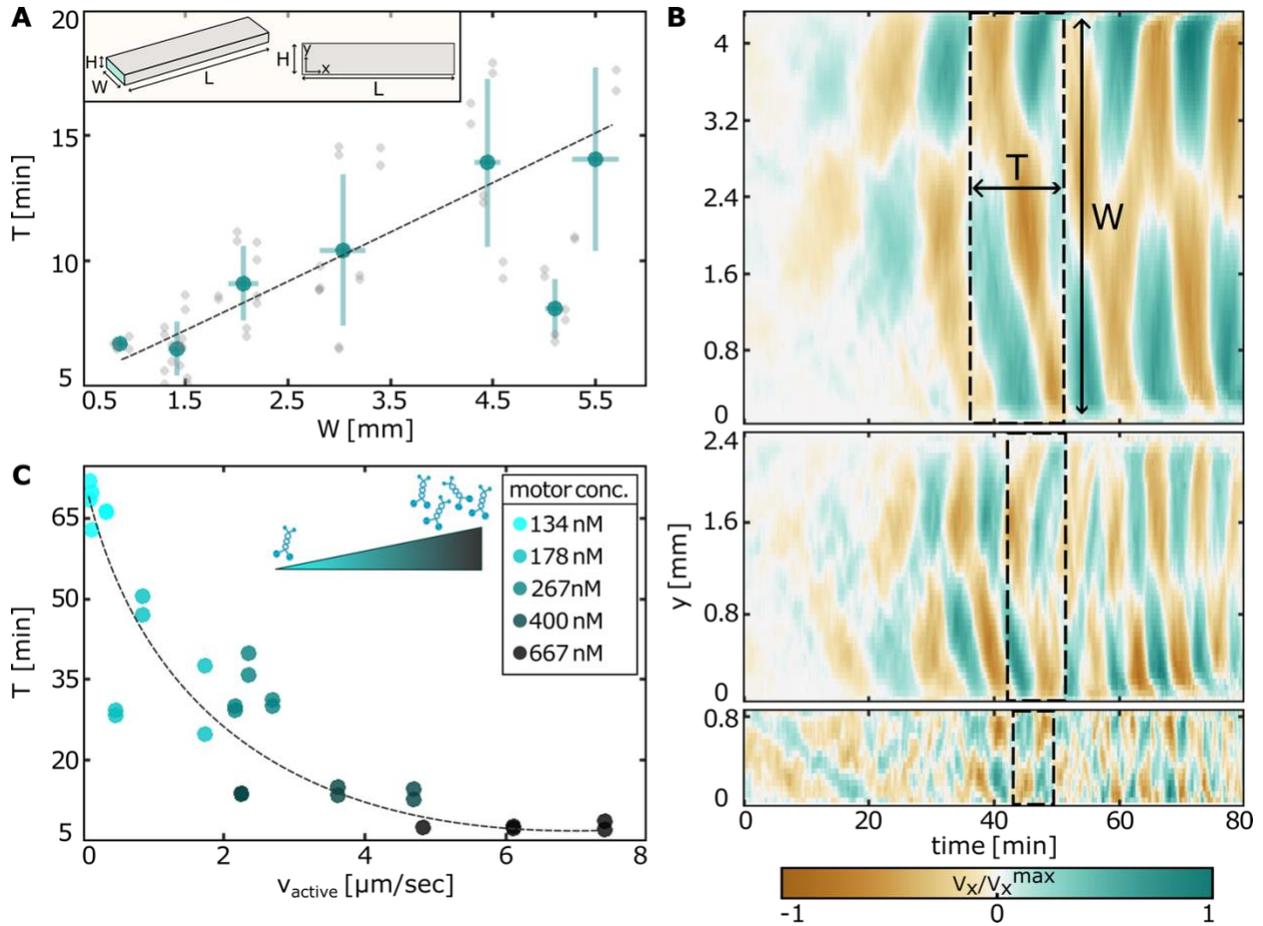

**Figure 4: Oscillation period scaling with geometry and activity.** (A) Oscillation period as a function of channel width $W$. Channel height, $H \approx 130 \, \mu m$, and length $L = 2cm$, were constant across all experiments. Horizontal error bars indicate standard deviation of the width over 3-5 experimental repeats. Vertical error bars indicate standard deviation of the oscillation period over 3-5 repeats. Single experiment results of the period of oscillations along $x$ and $y$ are presented using gray markers. (B) Space-time plots of $v_x$ along the short axis of the channel, y, for different channel widths. Dashed box indicates one oscillation period. Complementary space-time plots of $v_y$ are available in Fig. S15. (C) Oscillation period as a function of active velocity, controlled by kinesin motor concentration (indicated in the legend). For consistency each graph represents a distinct data set.

**Supplementary Information for:**

**This PDF file includes:**

**Methods**

**Supplementary text**

**SI references**

**Figures S1-S19**

**Video captions**

## Methods

**Tubulin purification and labeling:** Tubulin was purified from bovine brain by two successive cycles of polymerization and depolymerization as described previously (*1–5*). Briefly, polymerization was performed in M2B buffer (80 mM K-PIPES, 2 mM $MgCl_2$, 1 mM EGTA; pH 6.8) supplemented with Guanosine triphosphate (GTP, 100 mM; Aaron Chemicals, AR00I7U9). Following purification, tubulin was flash-frozen in liquid nitrogen and stored at -80 °C.

For subsequent use, purified tubulin was either recycled through one additional polymerization/depolymerization cycle or labeled with Alexa Fluor 647 NHS ester (Thermo Fisher, A20006) at a 13-fold molar excess of dye relative to tubulin. Recycled and labeled tubulin were flash-frozen in liquid nitrogen and stored at -80 °C.

**Microtubules polymerization:** Microtubules (MTs) were polymerized from purified tubulin at 8 mg $mL^{-1}$ in M2B buffer supplemented with 20 mM dithreitol (DTT) and 10 mM guanosine-5′-(α,β-methyleno)triphosphate (GMPCPP; Jena Bioscience), a non-hydrolysable GTP analogue. Reactions were incubated at 37 °C for 30 min to initiate polymerization and then at room temperature (RT) for 5 h to allow growth. Polymerized MTs were flash-frozen in liquid nitrogen and stored at -80 °C. For labeled MTs, 2-5 mol% Alexa Fluor 647 labeled tubulin was mixed with unlabeled tubulin prior to polymerization (*1*).

**Kinesin purification:** A dimeric kinesin construct comprising the N-terminal motor domain of *Drosophila melanogaster* kinesin-1 (residues 1-401) fused to the 87-amino acid Biotin Carboxyl Carrier Protein (BCCP) from *Escherichia coli* acetyl-CoA carboxylase (K401-BCCP, Addgene plasmid #15960) was expressed in the presence of biotin (24 mg $mL^{-1}$; Sigma-Aldrich, B4639) and purified using nickel-affinity chromatography, then dialyzed into M2B buffer supplemented with Adenosine triphosphate (ATP, 10 $\mu M$). Aliquots were flash-frozen in liquid nitrogen and stored at -80 °C (*1, 6*).

**Kinesin-streptavidin motor clusters:** Purified biotinylated kinesin was clustered with tetrameric streptavidin (*1*) (SA; Sigma-Aldrich, S4762). SA and kinesin were mixed at a 1:2 (v/v) ratio (SA:kinesin) in M2B containing 1 mM DTT, yielding final concentrations of ~0.47 mg $mL^{-1}$ kinesin and 0.12 mg $mL^{-1}$ SA. The mixture was incubated on ice for 30 min to allow cluster formation and flash frozen in liquid nitrogen (*1*).

**λ DNA concatenation**: λ DNA (NEB, N3011L, bacteriophage lambda cI857ind1 Sam7) was concatenated by ligating cohesive ends with T4 DNA ligase (NEB, M0202L, 10,000 units/ml). Reactions contained 350 ng/μl DNA (~11.7 nM), T4 DNA ligase (M0202L), in 1x ligation buffer (NEB B0202SVIAL, 50 mM Tris-HCl, 10 mM $MgCl_2$, 1 mM ATP, 10 mM DTT, pH 7.5@25˚C). To control the length distribution, a CY-5 labeled capping oligonucleotide (IDT, HPLC purified) was added at concentrations range 0-117 nM. Sequence with internal modification: GGGCGGCGACCT-Cy5-GGGTTGTCTCCTCTGATAGC. Ligation mixes were incubated in an Eppendorf Master cycler X50: 15 min at 25 °C, 10 h at 15 °C, then enzyme deactivation for 20 min at 65 °C. Products were kept at room temperature and used within 1 day. Before use, the ligated DNA samples were mixed by pipetting at least 30 times to improve solution homogeneity. To avoid degradation of the cohesive ends over repeated thawing and freezing rounds, the λ DNA solution was mixed by pipetting at least 50 times using a 1000 μL tip, aliquoted to working volume of 10 μL and kept at -20 ˚C. Before use, the DNA was mixed by pipetting and the concentration was measured using a spectrophotometer (DeNovix, DS-11FX). If needed, DNA volume in the concatenation reaction was adjusted to ensure a DNA concentration of 350 ng/µl.

**Gel electrophoresis and DNA length distribution analysis**: Concatenated DNA lengths were estimated by pulsed-field gel electrophoresis (PFGE) on 1% agarose (Sigma-Aldrich, A2929 or Lonza, SeaKem LE Agarose 50004) in 1x TAE buffer (40 mM Tris, 20 mM acetic acid, 1 mM EDTA) (*7*). For each DNA solution, three replicates were loaded (0.5 µL; 175 ng) mixed with 0.5x loading dye (Thermo Scientific). A λPFG ladder (NEB, N0341S) served as the size standard. Gels were run on a CHEF-DR III at 5.5 V cm$^{-1}$, 14 °C for 20 h with a ramp from 1-90s. After electrophoresis, gels were stained in EtBr/0.25x TAE for 60 min and imaged on a Typhoon FLA 9500 (532 nm and 647nm) or UVITEC Essential V6. Band fluorescence was used to estimate size distribution (band intensity normalized to total lane intensity). Reported values are mean ± standard deviation across the three replicates.

**λ DNA concentration scan:** Commercial λ-DNA (500 ng/µL) was concentrated by centrifugal ultrafiltration (Amicon Ultra, 50 kDa MWCO) prior to mixing with the active gel to reach target concentrations >225 ng/µL. Samples were spun at 1,000-2,200 x g (up to ~4 h), with mixing and concentration measurements using a spectrophotometer (DeNovix, DS-11FX) between spins. Filtration was continued until the desired concentration was reached. The final retentate volume was not predetermined and varied by sample. Integrity was verified by PFGE of a small aliquot alongside the unconcentrated stock to confirm no mechanical damage was done due to shear.

**Poly-ethylen glycol (PEG) stock solution:** A 10% (w/v) PEG-100 kDa (Sigma-Aldrich, 181986) stock was prepared in M2B buffer by vortex. PEG did not fully dissolve and formed visible aggregates. The stock was stored at RT and vortexed at maximum speed for at least 5 minutes before use. If needed, PEG aggregates were resuspended by pipetting with a 1000 µL tip, before proceeding with vortex mixing.

**Polyacrylamide-coated glass slides coating**

Glass slides (No. 1.5) were cleaned by boiling in ethanol for 5-10 min, rinsing three times in DDW, then boiling in 3% (v/v) Hellmanex solution for 10 min and rinsing three times in DDW. After cleaning, slides were dried and immersed for 15 min in a freshly prepared silane-coupling solution (1% acetic acid and 0.5% 3-(trimethoxysilyl)propyl methacrylate (Sigma-Aldrich, 440159), in ethanol), then rinsed three times in DDW and dried. To form the polyacrylamide brush, slides were placed in a 2% acrylamide solution (diluted from 40% acrylamide solution, (Bio-Rad, 1610140) degassed after dilution) supplemented with 0.035% (v/v) N,N,N′,N′-tetramethylethylenediamine (Sigma-Aldrich, T9281)) and 0.07% (w/v) ammonium persulfate (Sigma-Aldrich, A3678). Slides were stored in the acrylamide polymerization solution at room temperature for up to 3 months, and for at least 5 days before use. Immediately before use, slides were rinsed with DDW.

**Flow channel fabrication:** Channels were built from parafilm (*1*, *8*). A Parafilm spacer was cut, using silhouette 2.0, to the desired geometry and was placed between two polyacrylamide-coated slides, warmed to ~70 °C, and the top slide was gently pressed to seal.

**Microtubules-DNA composite active mix:** Concatenated λ-DNA was mixed with active fluid containing MTs (1.6 mg mL$^{-1}$), ATP (1.667 mM), SYBR Green I (1x), kinesin-streptavidin motor clusters (267 nM kinesin), an antioxidant cocktail (Trolox 2.1 mM; glucose 3.4 mg mL$^{-1}$; DTT 5.7 mM; glucose oxidase 230 µg mL$^{-1}$; catalase 40 µg mL$^{-1}$), an ATP-regeneration system (PEP 27.6 mM; PK/LDH 2.9% v/v) in M2B(*1*, *4*, *9*). Depending on the experiment, KSA and DNA concentrations were varied, unless noted otherwise, DNA was 70 ng μL$^{-1}$.

**Sample reproducibility:** To improve sample reproducibility, several steps were taken:

- To avoid degradation of the cohesive ends over repeated thawing and freezing rounds, the λ DNA solutions was mixed by pipetting at least 50 times using a 1000 μL tip and aliquoted to working volume (10 μL). Before use, the DNA was mixed by pipetting and the concentration was measured. If needed, DNA volume in the concatenation reaction was adjusted to ensure a DNA concentration of 350 ng/ul.

- Ligated DNA samples were mixed by pipetting at least >30 times before use, to improve solution homogeneity.

- Samples were highly sensitive to PEG conditions. Prior to use, the 100 kDa PEG solution was gently mixed with a 1000 μL pipette tip to resuspend aggregates, followed by vigorous vortexing at top speed for over 5 minutes.

**Tracer particles preparation:** For particle tracking, 2 μm fluorescent latex beads (Sigma Aldrich, L4530) were passivated with PLL-PEG coating: 0.2 mg mL$^{-1}$ PLL-PEG (SuSoS, PLL(20)-g[3.5]-PEG(2)) was dissolved in 10 mM HEPES and stored at 4°C for 1-2 weeks. Beads were diluted to 1.2% (v/v) in a 1% (v/v) PLL-PEG solution and incubated on a rotating platform for 60 min, mixed every 20 minutes by brief sonication (2-3 seconds, 3x. Sonication must be brief to avoid damaging the beads) (*10*). After 60 minutes, beads were washed by centrifuging (5140 xg, 5 min) twice and resuspend in 10 mM HEPES (final stock concentration 2% v/v). Finally, beads were added to the active-elastic composite at 6.5% of the stock solution (v/v). For these assays, DNA labeling with SYBR Green was removed to avoid spectral overlap with the bead fluorescence.

**Data Acquisition**: Fluorescence time-lapse imaging was performed on a Nikon Ti2 microscope with ORCA-Fusion CMOS camera (C14440-20UP/C15440-20UP). Large areas were acquired as multi-position tiles and stitched in FIJI (Grid/Collection Stitching).

- **Wide field:** MT and DNA were imaged with a 4x/0.20 NA objective (Nikon CFI Plan Apo 4x). High-magnification datasets used a 25x Si-oil/1.05 NA objective (CFI Plan). Frame interval: 3-8 seconds.

- **3D volumes:** Spinning-disk confocal (Crest Optics X-Light V3), 25x Si-oil/1.05 NA objective (Nikon CFI Plan Apo Lambda S 25XC Sil). Frame interval between z-stacks: 3 seconds (Fig. 2C) or 60 seconds (Fig. 2A, B).

- **Particle tracking:** Single-plane spinning-disk with 10x/0.45 NA (CFI Plan Apo Lambda D 10x). Frame interval 6 seconds. The focal plane was set away from the surfaces to avoid surface-adhered beads.

**Data Analysis:** Velocity fields were obtained by particle image velocimetry (PIV) on MT/DNA fluorescence time-lapses using PIVlab (MATLAB) using 16x16-52x52 μm$^2$ bins. For each time step, a velocity matrix was computed from consecutive frames. PIV results from the MT and DNA channels were consistent at low magnifications (<10x), showing no significant differences within measurement error (Fig. S6). Analyses shown were performed on MT fluorescence time-lapses.

**Maximum speed estimate:** For every PIV interrogation region we computed the temporal speed trace and applied a median filter to suppress outliers. The sample maximum velocity magnitude was defined as the 95th percentile of all filtered speed values pooled over time and space.

**3D datasets:** For spinning-disk confocal image stacks, we generated a maximum-intensity z-projection at each time point and applied the same PIV pipeline to the projected images.

**Supplementary Text**

**Correlation length and average velocity magnitude analysis:** The velocity-velocity correlation length, $\xi_v$, was estimated by calculating the velocity-velocity correlation function $C_{vv}$. From PIV fields $\vec{v}(\vec{r}, t)$, we computed the orientation and space-averaged, normalized correlation

$$C_{vv}(\vec{r}, t) = \langle \frac{\vec{v}(\vec{r}_0, t) \cdot \vec{v}(\vec{r}_0 + \vec{r}, t)}{v^2(\vec{r}_0, t)} \rangle_{\vec{r}_0, \theta}$$

such that $C_{vv}(\vec{r} = 0, t) = 1$. Within a single vortex, the velocity-velocity correlation function, $C_{vv}(\vec{r}, t)$ is positive at small separations due to co-rotation and becomes negative at larger separations corresponding to opposite, counter-rotating sides. We treat the entire vortex as a single correlated structure and fit only the initial decay of $C_{vv}(\vec{r}, t)$ from $r = 0$ to the first minimum using an exponential using an exponential function with an offset, allowing for negative values:

$$C_{vv}(\vec{r}, t) = a \cdot \exp\left(-\frac{\vec{r}}{\xi_v}\right) - C_0,$$

where $\xi_v$ is the correlation length and $C_0 \in [0, 1.2]$ allows negative values. Before fitting, we performed a moving average over $r$ to smooth the function, and a median filter (Fig. S17). The fit is restricted to the monotonic decay region in $C_{vv}(\vec{r}, t)$, and accepted for $R^2 > 0.9$.

For consistency, we analyzed the non-normalized covariance $\widetilde{C_{vv}}(\vec{r}, t)$ and extracted the spatial average of the velocity magnitude by calculating the non-normalized velocity-velocity correlation function:

$$\widetilde{C_{vv}}(\vec{r}, t) = \langle \vec{v}(\vec{r}_0, t) \cdot \vec{v}(\vec{r}_0 + \vec{r}, t) \rangle_{\vec{r}_0, \theta},$$

and fitting it as

$$\widetilde{C_{vv}}(\vec{r}, t) = v \cdot \exp\left(-\frac{\vec{r}}{\xi_v}\right) - C.$$

We report the mean speed as $v$.

**Oscillation Period analysis:** We defined the oscillation interval as the time window beginning at the onset of sustained amplitude growth, following initial increase in amplitude and ending at the decay of oscillatory behavior. Within this window, we analyzed $v_x(t)$ and $v_y(t)$ separately for each PIV interrogation region. Time traces were smoothed using a 30-second moving average and subsequently detrended by removing the DC component. The power spectrum was computed using the Fast Fourier Transform (FFT), and dominant frequency was identified from the maximal peak; if two nearby peaks of similar power appeared, we chose the higher frequency, and if two well-separated peaks of comparable power were present, the window was discarded. The resulting spatial map of frequencies was then averaged across windows to obtain the sample frequency along each axis. The two axes yielded similar values

**Spatial kinetic energy spectrum:** To analyze the scale-dependent distribution of kinetic energy, we computed the two-dimensional (2D) power spectrum of the instantaneous velocity field. Velocity magnitudes $|v(\vec{r})|$ were obtained from PIV measurements, and the Fast Fourier Transform (FFT) was applied to each frame to compute $|v(k)|$ in Fourier space. The power spectrum was then calculated as:

$$E(\vec{k}) = \frac{L_x L_y}{2} |v(\vec{k})|^2,$$

where $L_x$ and $L_y$ are the dimensions of the field of view. To obtain the one-dimensional radial energy spectrum, we averaged $E(\vec{k})$ over the angular coordinate $\theta_k$ in Fourier space and multiplied by $k$ to account for radial weighting. The resulting spectra were averaged over 10-minute time intervals to increase the statistics. To facilitate comparison, each spectrum was normalized by the energy of the lowest-energy mode, $E_{min}$.

**Particle tracking:** Trajectories and displacement fields were obtained by particle tracking velocimetry (PTV) of fluorescent tracer beads (*1*). Particles were segmented using size and fluorescence-intensity thresholds, then linked frame-to-frame to yield positions $\vec{x}(t)$ and displacements $\Delta\vec{x}$

**Intensity variance:** We quantified DNA network dynamics by measuring frame-by-frame spatial heterogeneity in the fluorescence intensity, $I(\vec{r}, t)$, of an intercalating DNA dye (Fig. 1). For each time point $t$, the intensity field was extracted from the fluorescence channel, and spatial intensity variance was computed over the full image domain. We quantified spatial heterogeneity using the normalized variance of the intensity:

$$\widetilde{\text{var}}(t) = \frac{\langle I^2 \rangle_{\vec{r}} - \langle I \rangle_{\vec{r}}^2}{\langle I \rangle_{\vec{r}}}.$$

where $\langle \cdot \rangle_{\vec{r}}$ denotes a spatial average over all pixels in the frame at time $t$. This metric captures second-order spatial fluctuations in signal intensity while minimizing sensitivity to uniform shifts in intensity caused by photobleaching, drift in illumination, or dye concentration variability. Larger $\widetilde{\text{var}}$ correspond to enhanced spatial heterogeneity and are indicative of network coarsening or structure formation, while low values reflect homogenization or network dissolution.

**Network thickness estimation:** From 3D confocal volume we quantified the DNA network thickness over time (Fig. 2B). For each time point the field was tiled to 65x65μm² in plane regions. In each region we extracted the z-axial DNA fluorescent profile $I(z)$ and fitted to a Gaussian:

$$I(z) \sim \exp\left(\frac{(z-\mu)^2}{2\sigma^2}\right).$$

The local thickness was defined as $2\sigma$ and fits with $R^2 > 0.6$ were retained for each time point, and the estimated error is the standard deviation of $2\sigma$ values. To smooth the data, we performed a 15-minute moving average.

**Vortex trajectory detection:** To quantify vortex advection in the oscillatory regime, PIV velocity fields were temporally averaged over 1-4 minutes, reducing frame-to-frame noise. The vorticity was calculated as $\omega_{xy} = \frac{1}{2}\left(\frac{dv_y}{dx} - \frac{dv_x}{dy}\right)$. We then defined a normalized vorticity $\widetilde{\omega} = \omega_{xy}/|v|$, were $v$ is the local velocity. We identified candidate vortex cores as locations where the vorticity exceeded the 60th percentile of the field of view. To confirm these were vortices centers, potential cores had to show sign reversals of $v_x$ along the short axis y and of $v_y$ along the long axis x, consistent with circulation around a core. We visually inspected these locations and linked confirmed cores across frames by nearest-neighbor association to construct vortex trajectories (Fig. 3, S10).

**Theoretical model for the emergence of oscillations in a viscoelastic active fluid**

Our goal here is to discuss a minimal model that gives rise to the two results presented in the main text and discussed therein in relation to experimental observations, i.e., that the velocity correlation length $\xi_v$ increases significantly with the viscoelastic relaxation time $\tau_{ve}$, until it saturates at the geometric length (system size) $W$ and the subsequent emergence of large-scale oscillations, with a period $T$. It is important to note that we do not model directly the dependence on the experimentally controlled DNA length $L_{DNA}$. Instead, we assume that once it becomes comparable with the DNA-free velocity correlation length, a viscoelastic polymeric network forms. This infers a sharp increase in the viscoelastic to the strain alignment time scales ratio, $\tau_{VE}/\tau_R$, and an increase in the DNA contribution to the long-time viscosity $\eta$. We show below how this drastic change in the rheology of the system, modeled on the macroscopic level as an active nematic gel, can rationalize the experimental observations.

### 1. A two-dimensional active gel model

We model the suspension of MTs, kinesin motors, and DNA polymers as an active nematic gel(*11*, *12*). The theory is formulated at the (macroscopic, coarse-grained) hydrodynamic level, considering the long-time and large-scale evolution of three continuum fields: a viscoelastic stress tensor $\boldsymbol{\sigma}^{el}$ accounting for the deformation of the DNA polymers, the director field $\vec{n}$ of the MTs, and the center-of-mass velocity $\vec{v}$.

We focus on the 2D in-plane dynamics of 3D system in view of the system's geometry that satisfies $H \ll W, L$. As noted in the main text, this effective description gives rise to a hydrodynamic force density $-\Gamma\vec{v}$, where $\Gamma$ is an associated friction-like coefficient and $\vec{v}$ refers hereafter to the in-plane velocity, which corresponds to the experimentally measured velocity. The effective friction-like force density appears in the 2D force balance equation, which in the limit of low Reynolds numbers (i.e., when fluid inertia is negligible) takes the form

$$\Gamma\vec{v} = \boldsymbol{\nabla} \cdot \boldsymbol{\sigma}, \tag{1}$$

where $\boldsymbol{\sigma}$ is the total 2D (in-plane) stress tensor. For reasons to be made clear soon, we take the two-dimensional curl of Eq. (1), leading to

$$\partial_{xx}\sigma_{yx} - \partial_{yy}\sigma_{xy} + \partial_{xy}(\sigma_{yy} - \sigma_{xx}) - \Gamma(\partial_x v_y - \partial_y v_x) = 0 \tag{2}$$

in components representation (where $x$ and $y$ are the Cartesian coordinates). Eq. (2) is our basic force balance equation. Our next goal is to explain the different contributions to the total stress $\boldsymbol{\sigma}$, which will also clarify how it depends on the basic fields in the problem, i.e., $\boldsymbol{\sigma}^{el}$, $\vec{n}$ and $\vec{v}$.

We consider five contributions to $\boldsymbol{\sigma}$, which can be written as $\boldsymbol{\sigma} = \boldsymbol{\sigma}^{vis} + \boldsymbol{\sigma}^{a} + \boldsymbol{\sigma}^{lc} + \boldsymbol{\sigma}^{el} - P\boldsymbol{I}$. Here, $\boldsymbol{\sigma}^{vis}$ is a viscous stress emerging from the flow of the solvent and MTs (without the DNA), $\boldsymbol{\sigma}^{a}$ is the active stress emerging from the activity of kinesin motors (acting on the MTs bundles), $\boldsymbol{\sigma}^{lc}$ is the passive liquid-crystalline stress and $\boldsymbol{\sigma}^{el}$ is one of our fundamental fields discussed above, corresponding to the viscoelasticity associated with the DNA polymers. Finally, $P$ is the pressure that enforces incompressibility and $\boldsymbol{I}$ is the identity tensor. Note that the pressure is already eliminated when Eq. (2) is considered, due to applying the two-dimensional curl to Eq. (1). We discuss each one of the first three contributions separately. First, we introduce the strain-rate tensor, which can be expressed in components form as $v_{ij} = \frac{1}{2}(\partial_i v_j + \partial_j v_i)$ (it is also known as the rate-of-deformation tensor). Here and elsewhere, $i, j$ denote in-plane Cartesian coordinates, i.e., $i = x, y$ and $j = x, y$. The viscous stress $\boldsymbol{\sigma}^{vis}$ is then expressed in components form as $\sigma_{ij}^{\text{vis}} = 2\eta_0 v_{ij}$, where $\eta_0$ is the fluid viscosity due to the solvent and MTs (without the DNA). The active (extensile) nematic stress is given as $\sigma_{ij}^{\text{a}} = -\alpha n_i n_j$, with the magnitude $\alpha > 0$. To describe the passive liquid-crystalline stress tensor, we introduce the molecular vector field $\vec{h}$ (*13*), which can be expressed in components form as $h_i = h_\parallel n_i + K\nabla^2 n_i$, where $K$ is the Frank elastic constant within a one-constant approximation and $h_\parallel$ is a Lagrange multiplier that enforces $|\vec{n}| = 1$, assuming that the system is deep in the nematic phase. In terms of $\vec{h}$, the passive liquid-crystalline stress is $\sigma_{ij}^{\text{lc}} = \frac{1}{2}(n_i h_j - h_i n_j) + \frac{\nu}{2}(n_i h_j + h_i n_j)$. The first term is the antisymmetric stress due to rotations of the director, while the second is the contribution of shear alignment involving a coupling constant $\nu$. Taken together, the total stress can be expressed in components form as

$$\sigma_{ij} = 2\eta_0 v_{ij} - \alpha n_i n_j + \frac{1}{2}(n_i h_j - h_i n_j) + \frac{\nu}{2}(n_i h_j + h_i n_j) + \sigma_{ij}^{\text{el}} - P\delta_{ij}. \tag{3}$$

Note that in Eq. (3) we neglect higher-order nonlinear terms of the Ericksen stress that scale as $\partial_i n_k \partial_j n_k$ (*13*), where the summation convention of repeated indices is used hereafter. Using Eq. (3) inside Eq. (2), force balance is fully expressed in terms of $\vec{v}$, $\vec{n}$, and $\boldsymbol{\sigma}^{\text{el}}$. To close the model, we now specify the time evolution of the director $\vec{n}$ and the viscoelastic stress $\boldsymbol{\sigma}^{\text{el}}$. In the inertia-less limit considered here, the velocity $\vec{v}$ has no independent dynamics and is determined instantaneously by Eq. (2). The DNA polymers endow the gel with transient elasticity, so it behaves as a viscoelastic fluid. To linear order, the DNA's contribution follows a Maxwell model:

$$(1 + \tau_{\text{ve}}\partial_t)\sigma_{ij}^{\text{el}} = 2\eta v_{ij}, \tag{4}$$

with the characteristic relaxation time $\tau_{\text{VE}}$ and long-time viscosity $\eta$ that includes contributions from the DNA polymers and should therefore be distinguished from $\eta_0$. At short times $t \ll \tau_{\text{VE}}$, this model corresponds to an elastic solid with a shear modulus $G = \eta/\tau_{\text{VE}}$, whereas at long times $t \gg \tau_{\text{VE}}$, it describes a viscous fluid. Note that in Eq. (4) we restrict ourselves to an isotropic rheology characterized by a single relaxation time and viscosity. More involved, nonlinear rheologies can be considered (*14–17*).

The director field $\vec{n}$ describes the orientational order of the MTs and defines the principal axes of the active stress. We consider systems deep in the nematic phase, such that the director dynamics follow

$$D_t n_i = \frac{h_i}{\gamma} - \nu v_{ij} n_j - \frac{1}{\tau_{\text{R}}} \frac{\sigma_{ij}^{\text{el}}}{G} n_j. \tag{5}$$

The left-hand-side of Eq. (5) is the co-rotational derivative, $D_t n_i \equiv \partial_t n_i + v_j \partial_j n_i + \omega_{ij} n_j$, with the vorticity tensor $\omega_{ij} = \frac{1}{2}(\partial_i v_j - \partial_j v_i)$. The first term on the right-hand-side (RHS) describes rotational relaxation with rotational viscosity $\gamma$. The second term on the RHS corresponds to shear alignment. The last term represents strain–director coupling, characterized by the rotation time $\tau_{\text{R}}$, where $\sigma_{ij}^{\text{el}}/2G$ is the (visco)elastic strain tensor associated with the DNA polymers (further discussed below). This is the simplest, minimal coupling between the director and strain tensor, which suffices to explain the experimentally observed phenomena, as shown below. This is a dissipative coupling, either with the activity or with the viscoelastic stress, whose reciprocal term (coupling the rheology of the viscoelastic stress with the molecular field) is negligible (*18*). More general couplings have been proposed for active gels (*16, 19, 20*) and for active and passive nematic elastomers (*21, 22*).

Eq. (2), once Eq. (3) is substituted in it, together with Eqs. (4-5), constitute the model to be analyzed below.

## 1.1 A scalar representation

One can use incompressibility, expressed as $\nabla \cdot \vec{v} = 0$ (or alternatively $\sum_i v_{ii} = 0$, the strain-rate tensor is traceless), to further simplify the mathematical formulation. Specifically, to express the velocity field in terms of a scalar field, we introduce the stream function $\psi$, defined as $v_x = \partial_y \psi, v_y = -\partial_x \psi$. Eq. (4) implies that $\boldsymbol{\sigma}^{\text{el}}$ is also a traceless tensor, i.e., it is purely deviatoric. Under these conditions, within a linearized theory, one has $\boldsymbol{\sigma}^{\text{el}} = 2G\boldsymbol{\varepsilon}$, where $\boldsymbol{\varepsilon}$ is a deviatoric strain tensor describing the deformation associated with the DNA polymeric degrees of freedom (the polymeric network), expressed in components form as $\varepsilon_{ij} = \frac{1}{2}(\partial_i u_j + \partial_j u_i)$, where $\vec{u}$ is the displacement field of the polymeric network. The similarity with the traceless strain-rate tensor is evident, implying that we can likewise introduce the scalar field $\Psi$, defined as $u_x = \partial_y \Psi, u_y = -\partial_x \Psi$. It allows to express the elastic tensor field $\boldsymbol{\sigma}^{\text{el}}$ in terms of the scalar field $\Psi$.

In terms of $\psi$ and $\Psi$, Eq. (4) can be rewritten in scalar form as

$$(1 + \tau_{\text{VE}} \partial_t)\Psi = \tau_{\text{VE}} \psi. \tag{6}$$

Moreover, since the director $\vec{n}$ is a unit vector, we can express it in terms of the angle $\theta$ relative to the $x$ axis as $\vec{n} = \cos\theta\,\hat{x} + \sin\theta\,\hat{y}$ ($\hat{x}$ and $\hat{y}$ are Cartesian unit vectors in the $x$ and $y$ directions, respectively). With this representation, all appearances of $\vec{n}$ can be expressed in terms of the (scalar) angle $\theta$. Consequently, the model can be formulated in terms of the scalar fields $\psi$, $\theta$ and $\Psi$ instead of $\vec{v}$, $\vec{n}$ and $\boldsymbol{\sigma}^{\text{el}}$. This scalar representation is used next.

## 2. Linear stability analysis and additional simplifications

The equations above have a stationary and homogeneous steady state, given by $\theta = \theta_0$ and $\psi = \psi_0$. While this oriented quiescent state is not realized in the experiments, the analysis of the linearized equations around it is insightful with regards to the observed dynamics. This has been demonstrated in the theory of active turbulence, where the characteristic length scale of the linear nematic instability emerges as the typical vortex size in the chaotic, turbulent state (*23–25*).

Following a similar approach, we consider the homogeneous $\psi = \theta = \Psi = 0$ state and introduce small perturbations to the three fields $\vec{\chi} = (\psi, \theta, \Psi)$ of the form $\vec{\chi} = \delta\vec{\chi} \exp(i\vec{q} \cdot \vec{r} + \Omega t)$. Here, the components of $\delta\vec{\chi}$ are the small perturbation amplitudes, $\vec{q} = q(\cos\phi\,\hat{x} + \sin\phi\,\hat{y})$ is the wavevector, $q$ is the wavenumber, and the real (imaginary) parts of $\Omega$ describe the growth/decay rate (frequency). We linearize the equations with respect to $\delta\vec{\chi}$, leading to $\boldsymbol{M} \cdot \delta\vec{\chi} = 0$. A nontrivial solution for the latter exists when $\det(\boldsymbol{M}) = 0$, leading to the stability spectrum (also known as the "dispersion relation").

The full calculation of the stability spectrum is given in section 4 of the theory. Here, we focus on a simplified version, using $\eta \gg \eta_0$ and $\nu = 0$, where the former assumes that the active gel viscosity is dominated by the DNA contribution. Consistent with this assumption, experiments show that long DNA polymers cause a ~5-fold reduction in the mean velocity before a network forms (Fig. S8). This indicates that an unentangled DNA suspension leads to a measurable increase in viscosity, which is expected to rise substantially once a network forms, owing to the slow viscoelastic relaxation ($\eta = G\tau_{\text{ve}}$). We set $\nu = 0$ because DNA-free MT-kinesin active gels are governed by a bend instability. Taking $\nu = 0$ leaves this mechanism unchanged. In this limit, the director does not rotate under shear but simply rotates with the local vorticity (the $\omega_{ij}n_j$ term in the co-rotational derivative of Eq. (5)). Furthermore, considering Eq. (5) at long times $t \gg \tau_{\text{VE}}$, the ratio $\tau_{\text{VE}}/\tau_{\text{R}}$ plays a similar role to $\nu$ in the director dynamics. Therefore, we assume that the shear alignment at long times is dominated by $\tau_{\text{VE}}/\tau_{\text{R}}$.

Under these simplifying assumptions, we arrive at the quadratic equation

$$\tau_{\text{VE}}\Omega^2 = -\mu(q)\Omega - k(q), \tag{7}$$

with

$$\mu(q) = 1 + q^2\left[l_{\text{h}}^2 + \tau_{\text{VE}}\left(\frac{K}{\gamma} - \frac{\alpha}{2\Gamma}\cos 2\phi\right) + \tau_{\text{VE}}\frac{K}{4\Gamma}q^2\right] \tag{8}$$

$$k(q) = q^2\left[\frac{K}{\gamma}(1 + q^2 l_{\text{h}}^2) + \frac{1}{\Gamma}\left(\frac{\tau_{\text{VE}}}{\tau_{\text{R}}}\cos 2\phi - \frac{1}{2}\right)\left(\alpha\cos 2\phi - \frac{K}{2}q^2\right)\right], \tag{9}$$

where $l_{\text{h}} = \sqrt{\eta/\Gamma}$ is a hydrodynamic screening length. Equation (7) can be interpreted as the equation of motion of a damped harmonic oscillator, with an effective friction coefficient $\mu(q)$ and spring constant $k(q)$. Note that $\mu(q)$ is dimensionless and $k(q)$ is of inverse time dimension. The equation is solved by the following spectrum

$$\Omega(q) = -\frac{\mu(q)}{2\tau_{\text{VE}}} \pm \sqrt{\left[\frac{\mu(q)}{2\tau_{\text{VE}}}\right]^2 - \frac{k(q)}{\tau_{\text{VE}}}}. \tag{10}$$

Below we identify a parameter range in which the linear stability spectrum $\Omega(q)$ and its dependence on the viscoelastic relaxation time $\tau_{\text{VE}}$, give rise to distinct dynamic regimes consistent with the experimental observations. To be precise, we relate the experimental findings to two linear instabilities:

(i) **A stationary (non-oscillatory) instability at short viscoelastic times**: the system is unstable in the limit $\tau_{\text{VE}} \to 0$, where the spectrum is purely real and there exists a band of wavenumbers $Re[\Omega(q)] > 0$ and $Im[\Omega(q)] = 0$ over $0 < q < q_c$. The instability occurs for small wavenumbers and persists until a critical $q = q_c$, which we relate to the correlation length $\xi_v \sim 1/q_c$. We show below that the correlation length increases with the long-time viscosity $\eta$ and with the timescale ratio $\tau_{\text{VE}}/\tau_{\text{R}}$.

(ii) **An oscillatory (Hopf) instability at long relaxation times:** the system undergoes a Hopf bifurcation for sufficiently large relaxation times. This instability, which features $Re[\Omega(q)] > 0$ and $Im[\Omega(q)] \neq 0$ over some $q$'s range, requires transient elasticity and can be thought of as an "elastomeric instability"(19). Here, the range of wavenumbers for which $Re[\Omega(q)] > 0$ includes the smallest available wavenumber $q_0 \equiv \pi/W$ and the characteristic frequency of large-scale oscillations is set by $Im[\Omega(q_0)]$.

Our analysis below aims at identifying the linear instabilities and their associated spatial and temporal characteristics, and at comparing the resulting characteristic scales with experimental observations across the different dynamical regimes. We emphasize that the homogenous reference state is not realized in the experiments, and that the observed dynamics can not be fully captured by the linearized theory.

## 3. Analysis of the stability spectrum in view of the experimental observations

Next, we analyze the spectrum $\Omega(q)$ in Eq. (10), with $\mu(q)$ and $k(q)$ in Eqs. (8-9) respectively. To set the stage for the analysis, we begin by reviewing the DNA-free case, similar to that of active turbulence (*24, 25*). Next, we analyze the effect of DNA as $\tau_{VE}/\tau_R$ and $\eta$ increase. Finally, we discuss the oscillatory instability that occurs for larger relaxation times. As the relaxation time increases with the experimentally controlled polymer length, the analysis structure reflects the dynamical regimes in three scenarios with increasing characteristic polymer length.

### 3.1 The DNA-free, active turbulence regime

In the absence of DNA, the viscoelastic relaxation time vanishes, $\tau_{VE} \to 0$, such that $\Omega(q) = -k(q)/\mu(q)$ is real (i.e., $Im[\Omega(q)] = 0$). In this case, we have $\mu(q) = 1 + q^2 l_h^2 > 0$ for all wavenumbers $q$. A non-oscillatory instability would emerge once $k(q) = q^2 \left[\frac{K}{\gamma}\mu(q) - \frac{1}{2\Gamma}\left(\alpha \cos 2\phi - \frac{1}{2}Kq^2\right)\right] < 0$. It is evident that for large enough $q$, $k(q)$ is positive (recall that $K, \gamma, \Gamma$ and $\alpha$ are all positive parameters), such that the instability is limited to large length scales.

The leading-order expansion of $k(q)$ near $q = 0$ is quadratic and the pre-factor is negative for

$$\tilde{\Gamma} \equiv \frac{2\Gamma K}{\gamma \alpha} < \cos 2\phi \leq 1. \tag{11}$$

Note that the above-defined dimensionless friction $\tilde{\Gamma}$ can also be expressed as $\tilde{\Gamma} = 2\Gamma l_a^2/\gamma$ in terms of the active length $l_a \equiv \sqrt{K/\alpha}$. While isolated active nematics are always unstable on large scales (*11*), friction sets a threshold for instability, surpassed for sufficiently large activity $\alpha$ (*11*). The instability first emerges in the $\phi = 0$ direction, corresponding to a bend instability. The inequality in Eq. (11) is a necessary condition for the emergence of active turbulence in the DNA-free system. Since the latter is experimentally observed, we assume hereafter that this condition is satisfied.

The instability persists in a range of wavenumbers $0 \leq q \leq q_c$, where the critical wavevector is given by

$$q_c^2 l_a^2 = 2\frac{1 - \tilde{\Gamma}}{1 + 4\eta/\gamma}. \tag{12}$$

We evaluate the most unstable direction by setting the angle of the wavevector $\phi = 0$, which maximizes the destabilizing active term. This result can be generalized for different scenarios of momentum transfer in the $z$-direction (*25*).

Typical stability spectra for $\tau_{VE} \to 0$ in the stable and unstable cases are depicted in Fig. S18. The instability is characterized by two wavenumbers, which can generally be related to the velocity correlation length $\xi_v$: $q_c$ and the fastest growing mode, for which $Re[\Omega(q)]$ is maximal. In what follows, we identify the velocity correlation length as $\xi_v \sim 1/q_c$. This is motivated by results for frictionless active turbulent systems, where the fastest growing mode is $q = 0$ and $q_c$ is the only remaining length scale, as well as experiments on active nematics with external dissipation, where $1/q_c$ matches the measured correlation length (*25*). This indicates that the wavelength-selection mechanism is nonlinear.

### 3.2 The first instability: the correlation length grows with the viscoelastic timescale

Since DNA polymers introduce viscoelasticity, we examine how the spectrum $\Omega(q)$ in Eq. (10) depends on a finite and small $\tau_{VE}/\tau_R$. By analyzing Eqs. (8-9), we find that the linear instability picture discussed above for $\tau_{VE}/\tau_R = 0$ persists for finite small $\tau_{VE}/\tau_R$ values, where the range of instability $0 < q < q_c$ shrinks with increasing $\tau_{VE}/\tau_R$, while still having $Im[\Omega(q)] = 0$.

We calculate explicitly $q_c$ in Section 4 below and present here the result in a simplified form. We compare $q_c$ with the value $q'_c$ obtained in the limit $\tau_{VE} = 0$, with a finite viscosity $\eta'$ before a network is formed. Comparing the correlation lengths $q_c$ and $q'_c$ yields

$$\left(\frac{q'_c}{q_c}\right)^2 = \frac{\eta}{\eta'} f(\tau_{ve}/\tau_R), \tag{13}$$

where we have defined the function

$$f(x) = \begin{cases} \left(1 - \dfrac{2x}{1-\tilde{\Gamma}}\right)^{-1}, & x \leq 1/4 \\ \left(1 - \dfrac{1 - \frac{1}{8x}}{1-\tilde{\Gamma}}\right)^{-1}, & x > 1/4 \end{cases}. \tag{14}$$

Notably, the function $f(x)$ is monotnoically increasing and has a pole at a finite $x$. Here, increasing $\tau_{VE}/\tau_R$ drives a sharp decrease of $q_c$, inferring an increase of the instability length scale (*18*). The prefactor $\eta/\eta'$ captures the calming effect of the increased viscosity due to network formation (*16, 19*). As $\tau_{VE}/\tau_R$ and $\eta$ grow, $q_c$ eventually reaches $q_0$. Identifying $\xi_v \sim 1/q_c$, we conclude that the velocity correlation length rises until it saturates at the system size $\xi_v \sim W$, as discussed in the main text (Fig. S18).

In the experiments, the velocity correlation length sharply increases (by an order of magnitude) over a narrow range of DNA length $L_{DNA}$ values, from the active turbulence value $\xi_0 \ll W$ to the system size $W$ (Fig. 3C). As stated above, we interpret the growth in the correlation length to be the increase of $\tau_{VE}/\tau_R$ and $\eta$ once $L_{DNA}/\xi_0$ becomes of order unity, when a polymeric network forms. Upon further increasing $L_{DNA}$, oscillations emerge on a scale $\xi_v \sim W$ (Fig. 3C). As explained earlier, we are therefore interested in the properties of the spectrum $\Omega(q)$ as $\tau_{VE}$ is further increased.

### 3.3 The second instability: the existence of large-scale oscillations

As the stabilizing $\tau_{VE}/\tau_R$ and $\eta$ further increase, $k(q)$ becomes positive for all $q$'s accessible to the system, and the homogeneous state $\psi = \theta = \Psi = 0$ is stable. The range of $\tau_{VE}/\tau_R$ over which this holds depends on other parameters, notably the timescale ratio $\tau_0/\tau_R$, where $\tau_0(q_0) \equiv \Gamma/(\alpha q_0^2)$ is a friction-related active timescale, relevant for active dynamics on the system scale $\sim 1/q_0$ (note that $\tau_0(q_0)$ identifies with $\tau_a$ defined in the caption of Fig. S18 for $1/q_0 = l_a$). Viewing Eq. (7) as a damped harmonic oscillator, a positive spring constant $k(q)$ allows for oscillations given a sufficiently low friction coefficient (the spectrum has an imaginary part for $[\mu/(2\tau_{ve})]^2 - k/\tau_{VE} < 0$ ). The question is, therefore, how does the $\mu$ term vary with increasing $\tau_{VE}$?

Interestingly, as $\tau_{VE}$ increases, $\mu$ not only decreases, but can become negative, inferring an oscillatory instability (Hopf bifurcation). This instability is possible due to the transient elasticity of the system provided by the polymeric degrees of freedom and was referred to as an "elastomeric instability" in (*19*).

Moreover, $\mu$ becomes negative only if the coefficient of $\tau_{VE}q^2$ in Eq. (8) becomes negative, which exactly coincides with the condition for the emergence of active turbulence in Eq. (11), which is satisfied in our system.

The onset of the oscillatory instability (a Hopf bifurcation) is mathematically expressed as $\mu(q) = 0$ and $k(q) > 0$. It is evident from Eq. (8) that if $\mu(q)$ becomes negative, it first happens at a finite $q$. The onset of the oscillatory instability is demonstrated in Fig. S19. With increasing $\tau_{VE}$, more wavenumbers become unstable until at some point $q_0$ undergoes an oscillatory instability, with a frequency $\omega = Im(\Omega) = \sqrt{k(q_0)/\tau_{VE}}$. The value of $\tau_{VE} = \tau_c$ for which $\mu(q_0) = 0$ is given by

$$\frac{\tau_c}{\tau_0} = 4 \frac{1 + q_0^2 l_h^2}{2(1 - \tilde{\Gamma}) - q_0^2 l_a^2}. \tag{15}$$

It is important to note that we expect the hydrodynamic screening length $l_h = \sqrt{\eta/\Gamma}$ to increase as the polymeric network forms. In fact, as stated in the main text, we expect it to grow in this regime such that it is larger than the system size $W$ over the entire range the latter is varied on (Fig. 4). That is, in the oscillations regime we expect the following length scale separation $l_a \ll W \ll l_h$.

Using the condition $\mu(q_0) = 0$, we obtain an oscillatory instability on the system scale, with a frequency

$$\omega^2 = \frac{1}{\tau_0} \left[ \frac{K}{2\gamma} q_0^2 \left(1 - \frac{1}{2} q_0^2 l_a^2\right) + \left(\frac{1}{\tau_R} - \frac{1}{2\tau_{VE}}\right)\left(1 - \frac{1}{2} q_0^2 l_a^2\right) \right]. \tag{16}$$

We retain lowest-order terms in $q_0 l_a \ll 1$ and consider $\tau_{VE} \gg \tau_R$, assuming that we are well beyond the first instability ($k(q) > 0$ for all $q$). These yield

$$\omega \approx \frac{1}{\sqrt{\tau_0 \tau_R}} \sim \frac{1}{W}\sqrt{\frac{\alpha}{\Gamma \tau_R}}. \tag{17}$$

The instability is demonstrated in Fig. S19. The frequency above reveals a characteristic timescale (oscillations period) $T \sim W\sqrt{\Gamma \tau_R/\alpha}$. In the manuscript, we compare this timescale with the measured large-scale oscillations period at a fixed $L_{DNA}$, while varying system size $W$ and level of activity $\alpha$.

The origin of the oscillations is as follows: the extensile active stress, featuring the timescale $\tau_0$, deforms the polymer network and tends to align it with the director. The director, however, lags behind and rotates away due to strain alignment over the timescale $\tau_R$, providing an effective restoring force. Alternatively, this alignment behavior of the director and the network can be thought of as a non-reciprocal interaction, which can generically lead to oscillations (*26*). We note that even below the strict onset of the oscillatory instability, we have $|Re[\Omega(q_0)]| \ll Im[\Omega(q_0)]$ (with $Re[\Omega(q_0)] < 0$), which implies that decaying oscillations can persist for long times, and hence possibly be observed.

### 3.4 Summary of the analysis

The minimal and simple model we presented features two linear instabilities, whose onset depends on the viscoelastic timescale $\tau_{VE}$ (*18*). In the absence of DNA, $\tau_{VE} \to 0$, and with sufficiently large activity, a non-oscillatory instability gives rise to active turbulence (Eq. 11). In the presence of DNA at small $\tau_{VE}/\tau_R$ values, the instability persists. In this regime, increasing viscoelasticity stabilizes the system, leading to a decrease in $q_c$ and hence an increase in $\xi_v$. For sufficiently large $\tau_{VE}/\tau_R$ values, an oscillatory instability emerges, giving rise to a characteristic oscillation at the system size. Between these limits, the model predicts an intermediate linearly stable window, which may not be realized experimentally given the possible sharp increase in $\tau_{VE}$ as the DNA network forms (for $L_{DNA}$ comparable with the correlation length in the absence of DNA, $\xi_0$).

It is interesting that the same minimal and simple model reveals, in a unified manner, two instabilities as a function of $\tau_{VE}$, whose properties appear to be related to the much more complicated experimental system. Moreover, the necessary condition for the emergence of the two instabilities is the very same one (the inequality in (11), expressed in terms of $\tilde{\Gamma}$). The latter highlights the important role of the dimensionless combination of parameters lumped into $\tilde{\Gamma}$, revealing the interplay between activity ($\alpha$), effective friction ($\Gamma$), nematic elasticity ($K$) and rotational viscosity ($\gamma$). The dimensional ratio $\alpha/\Gamma$ also plays a central role in the emerging oscillations, as seen in Eq. (17).

## 4. Full derivation of the stability spectrum and analysis of the correlation length

### 4.1 Derivation of the stability spectrum

We consider small perturbations of the fields $\vec{\chi} = (\Psi, \theta, \psi)$ of the form $\vec{\chi} = \delta\vec{\chi} \exp(i\vec{q} \cdot \vec{r} + \Omega t)$, as described above. Linearizing Eq. (2) yields

$$(\Gamma + \eta_0 q^2)\psi_1 = -Gq^2 \Psi_1 + \left[\left(\alpha + \frac{\nu K}{2} q^2\right) \cos 2\phi - \frac{K}{2} q^2\right] \theta_1 + \frac{\nu}{2} h_\parallel \sin 2\phi, \tag{18}$$

while projecting Eq. (5) parallel and perpendicular to $n_\alpha$ yields

$$\left(\Omega + \frac{K}{\gamma} q^2\right) \theta_1 = \frac{q^2}{2}(1 - \nu \cos 2\phi)\psi_1 - \frac{q^2}{\tau_R} \cos 2\phi\, \Psi_1, \tag{19}$$

$$h_\parallel = -\frac{\gamma}{2} q^2 \sin 2\phi \left(\frac{2}{\tau_R} \Psi_1 + \nu \psi_1\right). \tag{20}$$

These equations can be written in matrix form as $\mathbf{M} \cdot \delta\vec{\chi} = \mathbf{0}$, where $\mathbf{M}$ is the dynamic matrix. A nontrivial solution for $\delta\vec{\chi}$ is obtained for $\det(\mathbf{M})=0$. This leads to the quadratic Eq. (7), with

$$\mu(q) = 1 + \frac{K}{\gamma}\tau_{VE} q^2 + \frac{q^2}{D}\left[\eta + \frac{\nu\gamma}{2} \cdot \frac{\tau_{VE}}{\tau_R} \sin^2 2\phi + \frac{C}{2}\tau_{VE}(\nu \cos 2\phi - 1)\right], \tag{21}$$

$$k(q) = \frac{Kq^2}{\gamma} + \frac{q^2}{D}\left[C\left(\frac{1}{2}(\nu \cos 2\phi - 1) + \frac{\tau_{VE}}{\tau_R}\cos 2\phi\right) + Kq^2\left(\frac{\eta}{\gamma} + \frac{\nu}{2}\frac{\tau_{VE}}{\tau_R}\sin^2 2\phi\right)\right], \tag{22}$$

where we have defined the functions

$$C = \alpha \cos 2\phi + \frac{K}{2} q^2(\nu \cos 2\phi - 1), \tag{23}$$

$$D = \Gamma + \left(\eta_0 + \frac{\gamma}{4}\nu^2 \sin^2 2\phi\right) q^2. \tag{24}$$

This reduces to Eqs. (8-9) by focusing on $\nu = 0$ and $\eta \gg \eta_0$, as discussed above. In this regime, the friction term dominates over the MT viscosity $\Gamma \gg \eta_0 q^2$, while the same does not hold for the long-time gel viscosity, allowing for large values of $l_h^2 = \eta/\Gamma$.

### 4.2 Analysis of the correlation length

The first instability relates to the scenario of $k(q) < 0$, $\mu(q) > 0$. It originates from activity, specifically a sufficiently large active stress term $\sim \alpha \cos 2\phi$, and persists for small wavenumbers $0 \leq q < q_c$. We solve for $q_c$ and find that

$$q_c = \frac{\cos 2\phi \left(\frac{1}{2} - \frac{\tau_{VE}}{\tau_R} \cos 2\phi\right) - \frac{1}{2}\tilde{\Gamma}}{\frac{1}{4} - \frac{1}{2}\frac{\tau_{VE}}{\tau_R} \cos 2\phi + R}. \tag{25}$$

Here, we have denoted the ratio between shear and rotational viscosities as $R = \eta/\gamma$.

As an angle-independent value of $q_c$, we set the angle $\phi$ of the wavevector such that it maximizes the destabilizing term in $k(q)$. This yields $\phi = 0$ for $\tau_{VE}/\tau_R < 1/4$ (related to pure bend instability) and $\cos 2\phi = \tau_R/4\tau_{VE}$ for $\tau_{VE}/\tau_R \geq 1/4$, combining bend and splay.

To better relate our results to experimental measurements, we compare the critical wavevector $q_c$ with the one obtained in the limit $\tau_{ve} = 0$, denoted by $q'_c$, as discussed above. We find that

$$\left(\frac{q'_c}{q_c}\right)^2 = \left[\frac{R + 1/4}{R' + 1/4} - \Delta(\tau_{VE}/\tau_R)\right] f(\tau_{VE}/\tau_R) \tag{26}$$

where we have defined the functions,

$$\Delta(x) = \begin{cases} \frac{x}{2R'+1/2}, & x \leq 1/4 \\ \frac{1/4}{2R'+1/2}, & x > 1/4 \end{cases} \qquad f(x) = \begin{cases} \left(1 - \frac{2x}{1-\tilde{\Gamma}}\right)^{-1}, & x \leq 1/4 \\ \left(1 - \frac{1-\frac{1}{8x}}{1-\tilde{\Gamma}}\right)^{-1}, & x > 1/4. \end{cases} \tag{27}$$

Notably, the function $f(x)$ has a pole, which implies a diverging critical wavelength. The value of $x = \tau_{VE}/\tau_R$ where this takes place depends on the value of $\tilde{\Gamma} < 1$ (the bound is implied by the emergence of active turbulence for $\tau_{VE} = 0$, Eq. (11)). For $\tilde{\Gamma} \geq 1/2$, we find a pole at $x = (1 - \tilde{\Gamma})/2$ and for $\tilde{\Gamma} \leq 1/2$, at $x = (8\tilde{\Gamma})^{-1}$.

The result of Eq. (26) is simplified, assuming that $R, R' \gg 1$, which is feasible for $\gamma \approx \eta_0$. In this case, $\Delta(x) \ll 1$ and $(R + 1/4)/(R' + 1/4) \approx R/R' \approx \eta/\eta'$. This yields the compact result in Eq. (13).

**References**

1. A. M. Tayar, L. M. Lemma, Z. Dogic, Assembling Microtubule-Based Active Matter. *Methods in Molecular Biology* **2430**, 151–183 (2022).

2. T. Sanchez, D. Welch, D. Nicastro, Z. Dogic, Cilia-like beating of active microtubule bundles. *Science (1979)* **333**, 456–459 (2011).

3. T. Sanchez, D. T. N. Chen, S. J. Decamp, M. Heymann, Z. Dogic, Spontaneous motion in hierarchically assembled active matter. *Nature* **491**, 431–434 (2012).

4. G. Henkin, S. J. DeCamp, D. T. N. Chen, T. Sanchez, Z. Dogic, Tunable dynamics of microtubule-based active isotropic gels. *Philosophical Transactions of the Royal Society A: Mathematical, Physical and Engineering Sciences.* **372** (2014).

5. M. Castoldi, A. V. Popov, Purification of brain tubulin through two cycles of polymerization-depolymerization in a high-molarity buffer. *Protein Expression and Purification* **32**, 83–88 (2003).

6. D. S. Martina, R. Fathib, T. J. Mitchisonc, J. Gellesa, FRET measurements of kinesin neck orientation reveal a structural basis for processivity and asymmetry. *Proceedings of the National Academy of Sciences USA* **107**, 5453–5458 (2010).

7. J. Herschleb, G. Ananiev, D. C. Schwartz, Pulsed-field gel electrophoresis. *Nature Protocols 2007 2:3* **2**, 677–684 (2007).

8. P. Chandrakar, M. Varghese, S. A. Aghvami, A. Baskaran, Z. Dogic, G. Duclos, Confinement Controls the Bend Instability of Three-Dimensional Active Liquid Crystals. *Physics Review Letters* **125**, (2020).

9. A. M. Tayar, M. F. Hagan, Z. Dogic, Active liquid crystals powered by force-sensing DNA-motor clusters. *Proceedings of the National Academy of Sciences* **118** (2021).

10. A. Azioune, M. Storch, M. Bornens, M. Théry, M. Piel, Simple and rapid process for single cell micro-patterning. *Lab Chip* **9**, 1640–1642 (2009).

11. M. C. Marchetti, J. F. Joanny, S. Ramaswamy, T. B. Liverpool, J. Prost, M. Rao, R. A. Simha, Hydrodynamics of soft active matter. *Review of Modern Physics* **85**, 1143 (2013).

12. J. Prost, F. Jülicher, J. F. Joanny, Active gel physics. *Nature Physics 2014 11:2* **11**, 111–117 (2015).

13. P.-G. de Gennes, J. Prost, The physics of liquid crystals (international series of monographs on physics). *Oxford University Press, USA*, 0–20 (1995).

14. S. T. Milner, Dynamical theory of concentration fluctuations in polymer solutions under shear. *Physics Review E* **48**, 3674 (1993).

15. H. Pleiner, D. Svenšek, H. R. Brand, Hydrodynamics of active polar systems in a (Visco)elastic background. *Rheologica Acta 2016 55:10* **55**, 857–870 (2016).

16. E. J. Hemingway, A. Maitra, S. Banerjee, M. C. Marchetti, S. Ramaswamy, S. M. Fielding, M. E. Cates, Active viscoelastic matter: From bacterial drag reduction to turbulent solids. *Physics Review Letters* **114**, (2015).

17. R. M. Adar, J. F. Joanny, Permeation Instabilities in Active Polar Gels. *Physics Review Letters* **127**, (2021).

18. S. Liu, S. Shankar, M. C. Marchetti, Y. Wu, Viscoelastic control of spatiotemporal order in bacterial active matter. *Nature* **590**, 80–84 (2021).

19. E. J. Hemingway, M. E. Cates, S. M. Fielding, Viscoelastic and elastomeric active matter: Linear instability and nonlinear dynamics. *Physics Review E* **93**, (2016).

20. S. Fürthauer, D. J. Needleman, M. J. Shelley, R. M. Adar, J.-F. , Ois Joanny, Active-gel theory for multicellular migration of polar cells in the extra-cellular matrix. *New Journal of Physics* **24**, (2022).

21. Mark. Warner, E. M. Terentjev, Liquid crystal elastomers. 407 (2007).

22. A. Maitra, S. Ramaswamy, Oriented Active Solids. *Physics Review Letters* **123**, (2019).

23. L. Giomi, Geometry and topology of Turbulence in active nematics. *Physics Review X* **5**, (2015).

24. R. Alert, J. F. Joanny, J. Casademunt, Universal scaling of active nematic turbulence. *Nature Physics* **16**, 682–688 (2020).

25. B. Martínez-Prat, R. Alert, F. Meng, J. Ignés-Mullol, J. F. Joanny, J. Casademunt, R. Golestanian, F. Sagués, Scaling Regimes of Active Turbulence with External Dissipation. *Physics Review X* **11**, 031065 (2021).

26. M. Fruchart, R. Hanai, P. B. Littlewood, V. Vitelli, Non-reciprocal phase transitions. *Nature 2021* **592**, 363–369 (2021).

**Supplementary Figures**

A 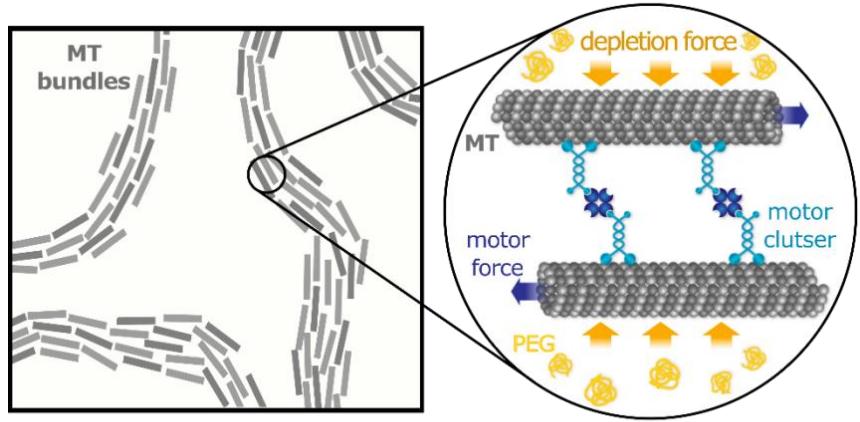 B 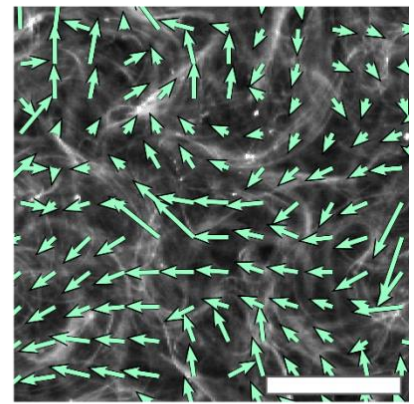

C 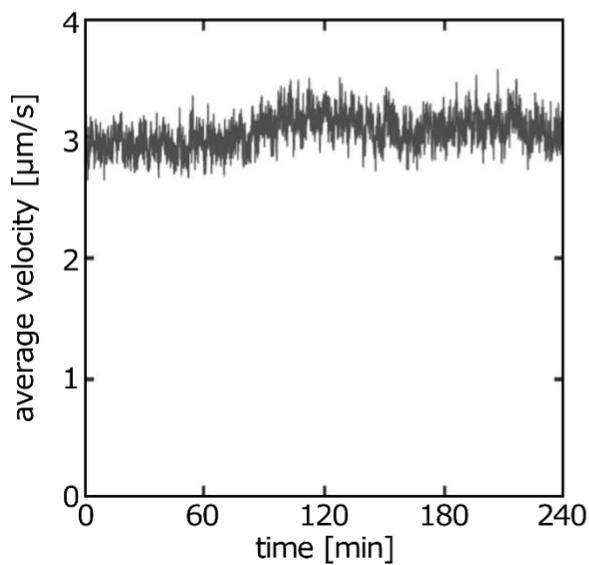 D 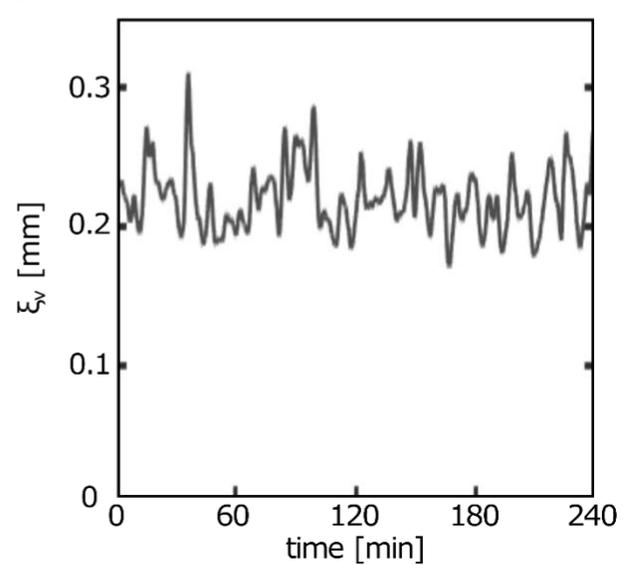

**Figure S1:** (A) Schematic of a DNA-free MT-kinesin active fluid. Left: PEG-bundled MT filaments. Right: kinesin-streptavidin motor clusters generating shear by sliding antiparallel MTs filaments. (B) Fluorescence image of an active MT fluid, overlaid with the PIV velocity field (Methods). Scalebar, 200 μ$m$. (C) Field average velocity magnitude within the field of view of 2 $mm$ x 3.8 $mm$ versus time. (D) Velocity-velocity correlation length $\xi_v$ versus time.

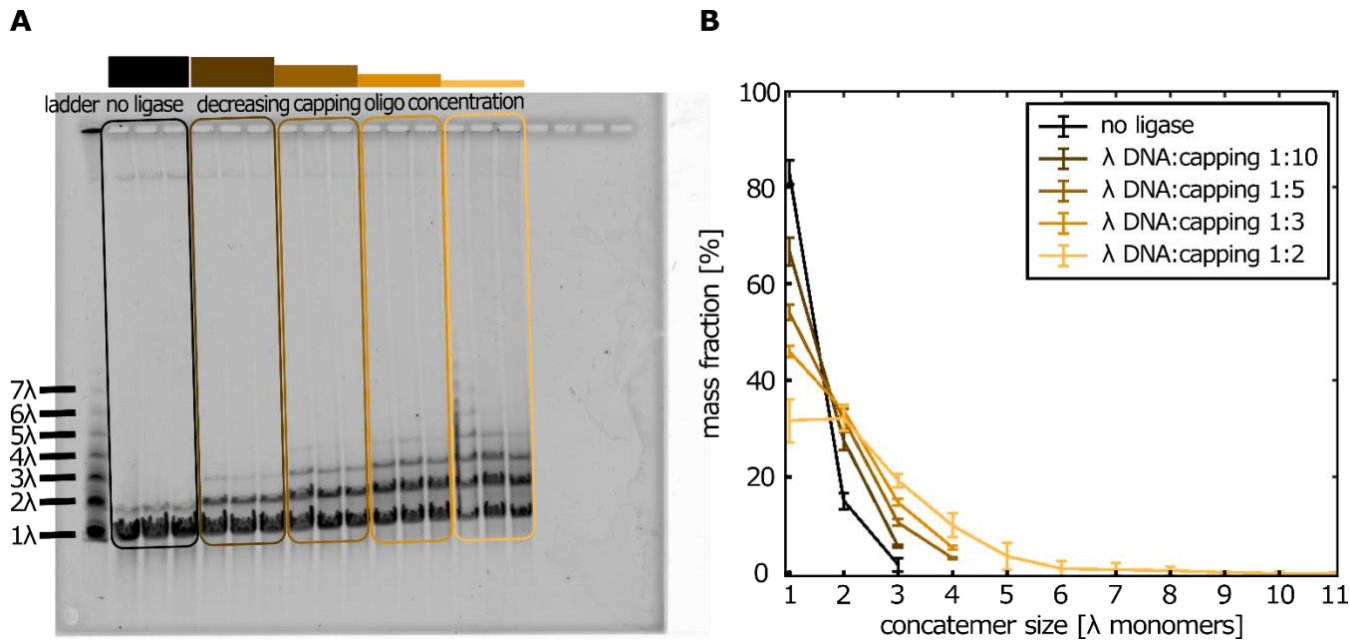

**Figure S2:** (A) Representative agarose DNA gel in PFGE of λ DNA samples concatenated under varying conditions. Lanes 1-16 from left to right: 1. Ladder (NEB). 2-4. λ DNA incubated at $C_{DNA} = 350 \ ng/\mu l$ without DNA ligase (negative control). 5-7, λ DNA concatenated at $350 \ ng/\mu l$ (~$17 \ nM$ λ DNA monomers) with capping oligo at a 10:1 (capping: DNA molar ratio). 8-10, 5:1 molar ratio. 11-13, 3:1 molar ratio. 14:16, 2:1 molar ratio. (B) Quantification of polymer length distributions by mass fraction at each length, based on fluorescence intensity normalized by the total intensity. Error bars represent the standard deviation across three replicate gel lanes per condition.

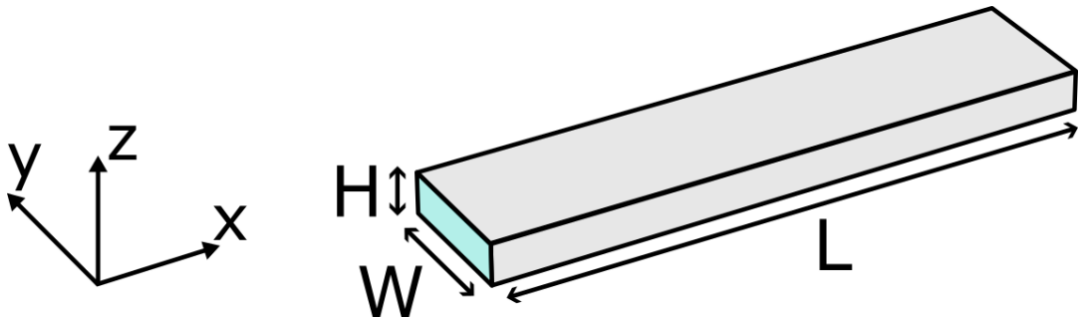

**Figure S3**: Schematic of the flow-channel geometry. Unless otherwise stated, the dimensions are $H = 0.13 \ mm$, $W = 2 \ mm$, $L = 22 \ mm$ such that $H \ll W \ll L$.

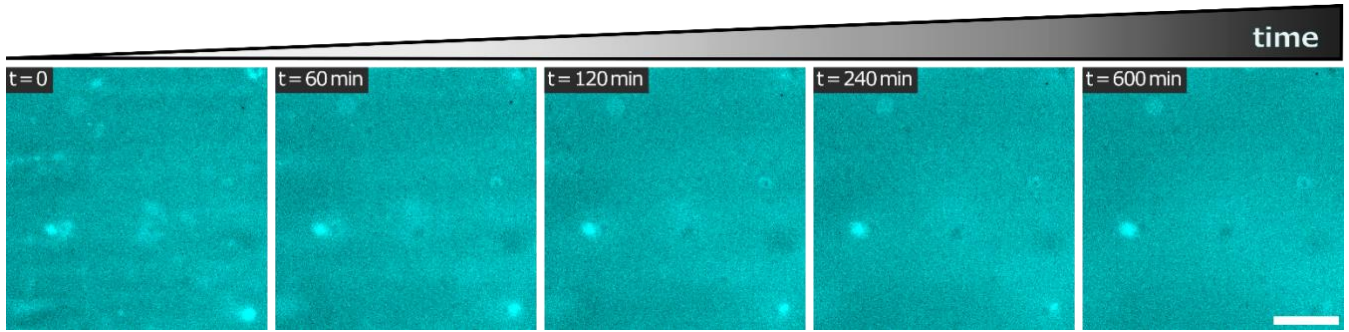

**Figure S4:** Time-lapse fluorescent images of concatenated λ-DNA labeled with SYBR Green at $C_{DNA} = 70\ ng/\mu l$ $L_{DNA} \approx 40\ \mu m$, in 1% PEG solution, in the absence of MT and kinesin motor clusters. Scale bar, 250 µm.

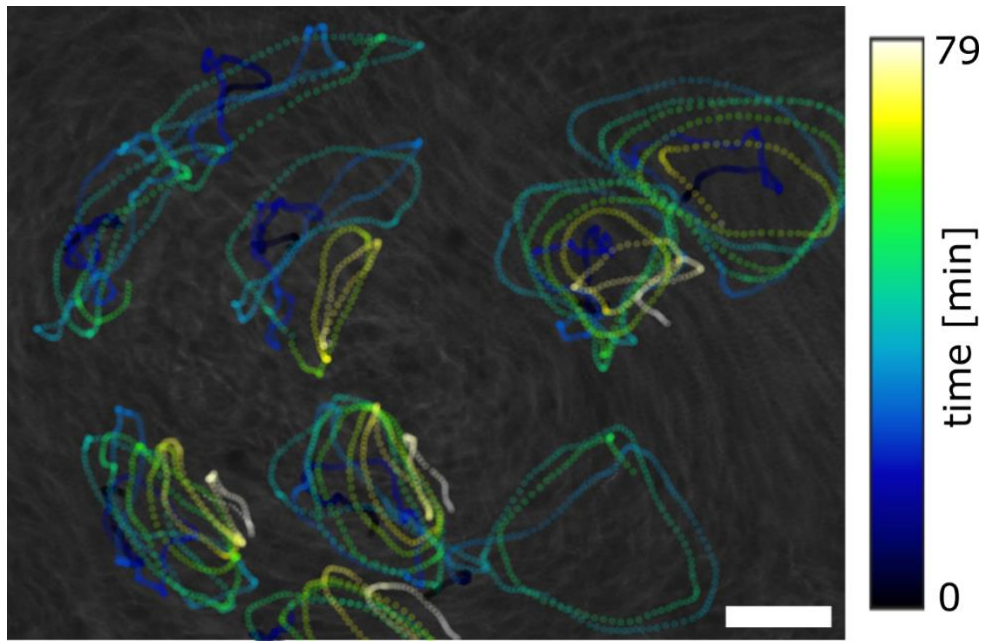

**Figure S5:** 80 min time-projection of 2 $\mu m$ tracer-particle trajectories from the experiment in Fig. 1G, overlaid on a 90-s time projection of MTs (gray). Scale bar, 250 $\mu m$.

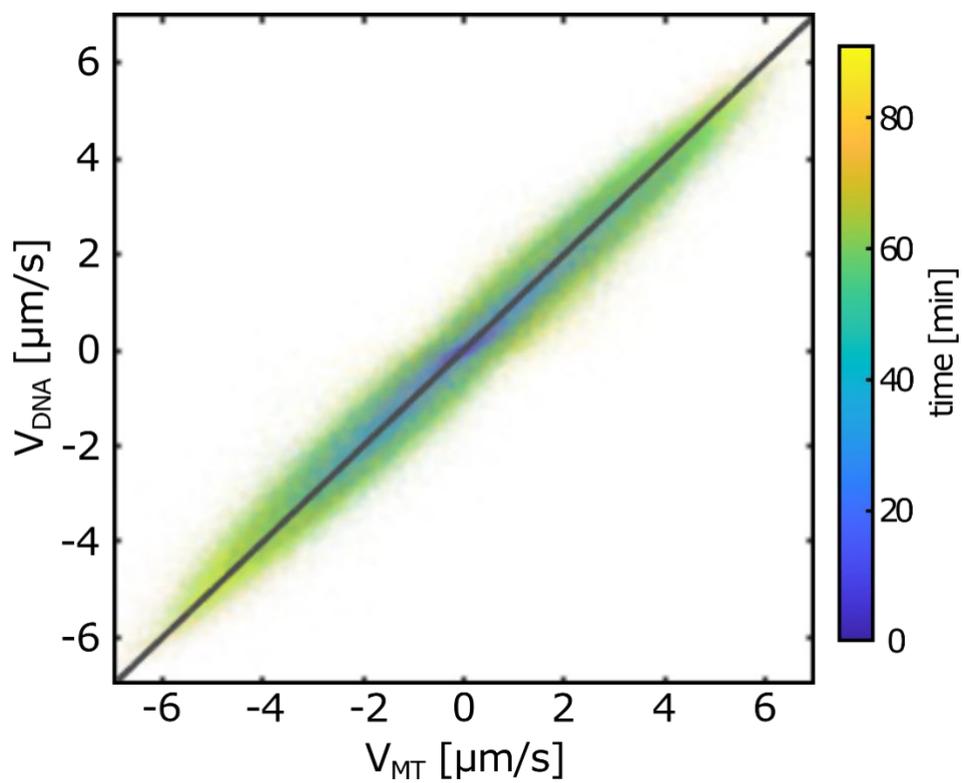

**Figure S6:** DNA velocity as a function of MT velocity. Gray line indicates a slope of 1. Each data point is the mean velocity within 67x67 µm² region of interest at one time point (color-coded). Both $v_x$ and $v_y$ components are plotted on the same graph.

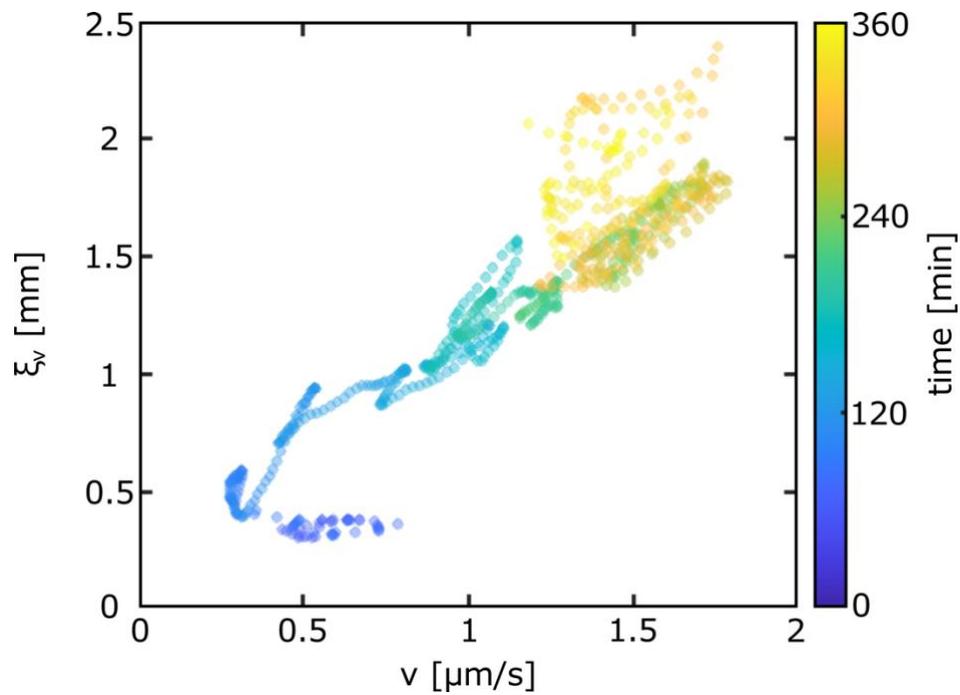

**Figure S7:** Velocity-velocity correlation length $\xi_v$ as a function of velocity magnitude over the entire field of view $\langle v \rangle$. Each point is a time point from the data shown in Fig. 2D. Color encodes experiment time.

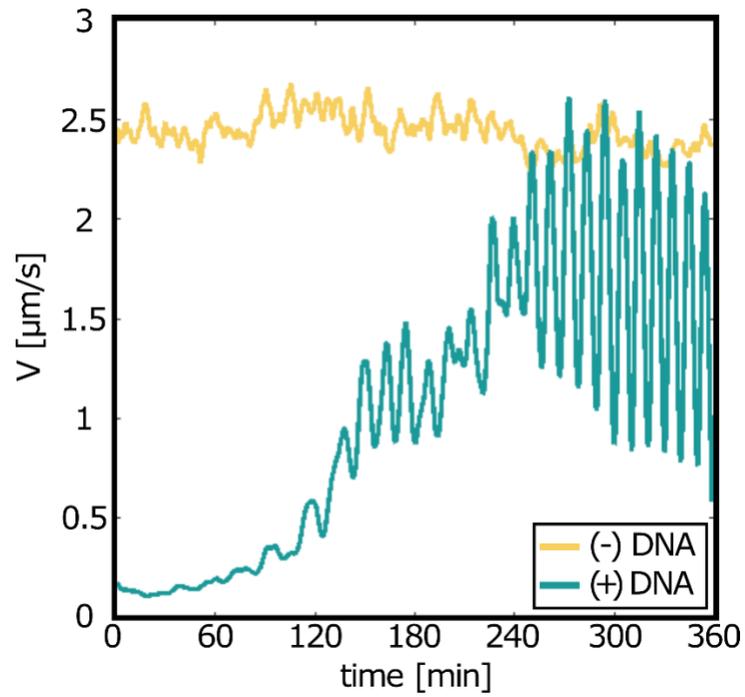

**Figure S8**: Mean velocity magnitude averaged over the field of view as a function of time, for active fluid samples with (+, green) and without (-, yellow) DNA. + DNA samples were at a concentration of $C_{DNA} = 70\ ng/\mu L$.

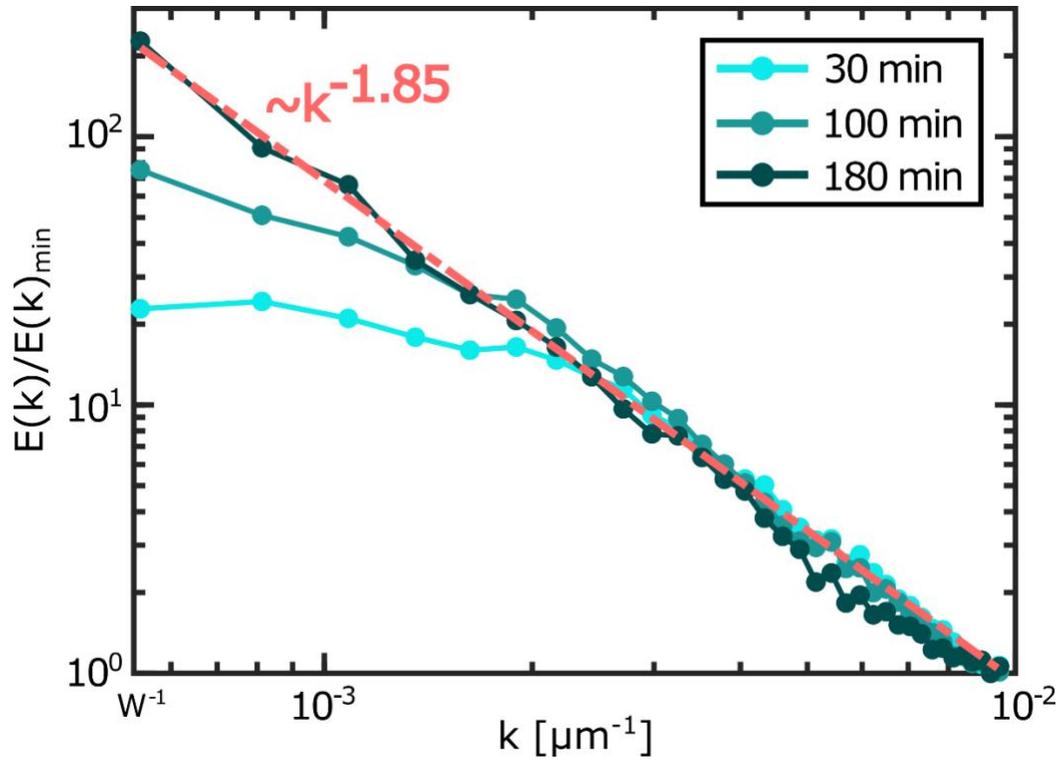

**Figure S9:** Normalized kinetic energy spectrum as a function of wavenumber $k$, at $t = 30, 100, 180 min$ (see legend). Spectra are computed from the velocity fields and normalized by each spectrum minimum kinetic energy. The confinement-imposed cutoff appears as $k \approx W^{-1}$, corresponding to the inverse channel width. Over an intermediate $k$-range the spectra follow a power law $E(k) \propto k^{-1.85}$.

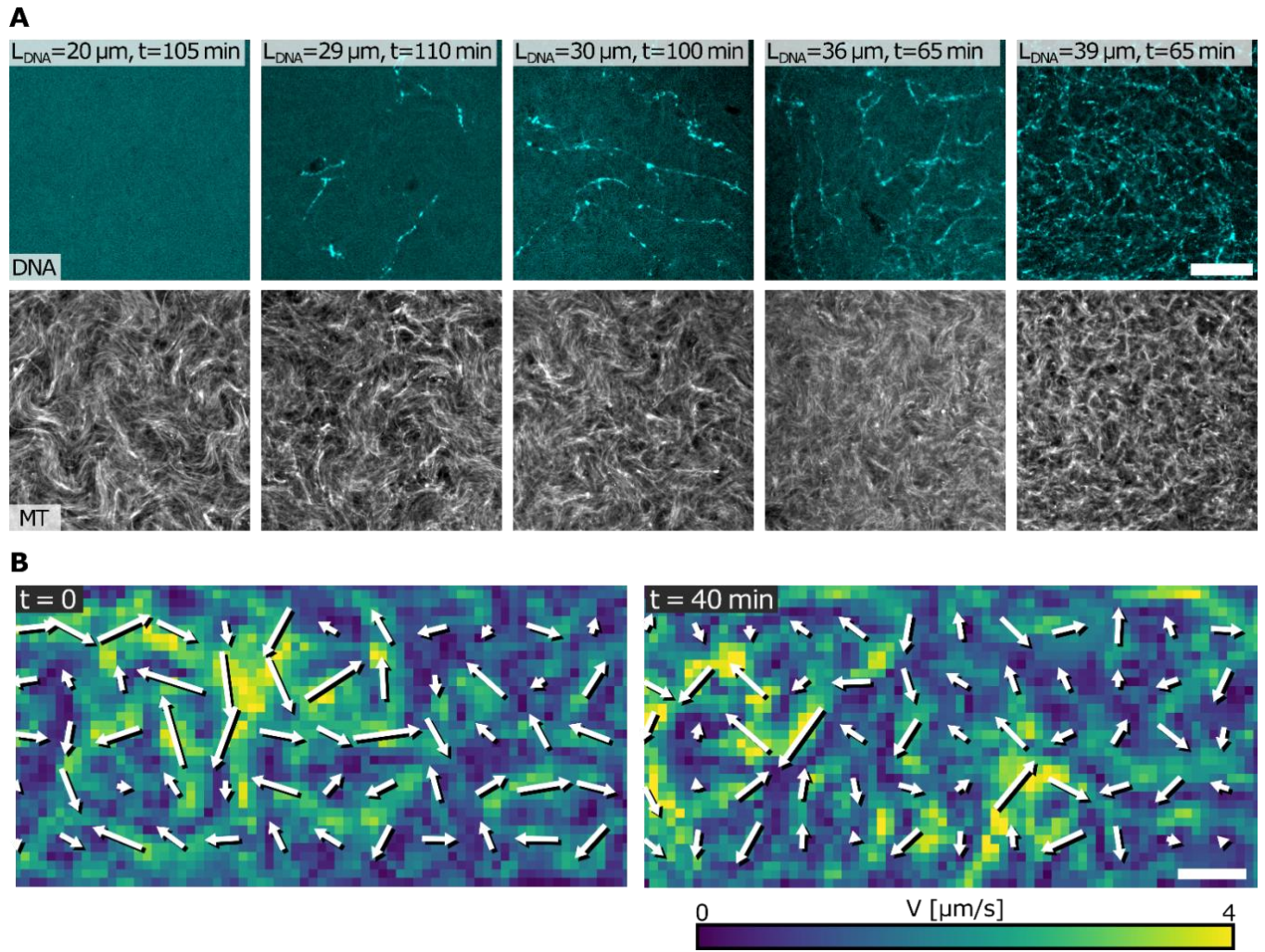

**Figure S10:** (A) Representative fluorescence microscopy micrographs showing DNA network organization at different average DNA contour lengths. The DNA concentration was fixed at 70 $ng/\mu l$. A simultaneous image shows MT (gray) organization at the same time point. Scale bar, 250 $\mu m$. (B) Velocity magnitude heatmaps and velocity fields of short DNA embedded in a MT active gel. The mean DNA contour length is $L_{DNA} \approx 27~\mu m$ and the DNA concentration is $C_{DNA} = 70~ng/\mu l$.

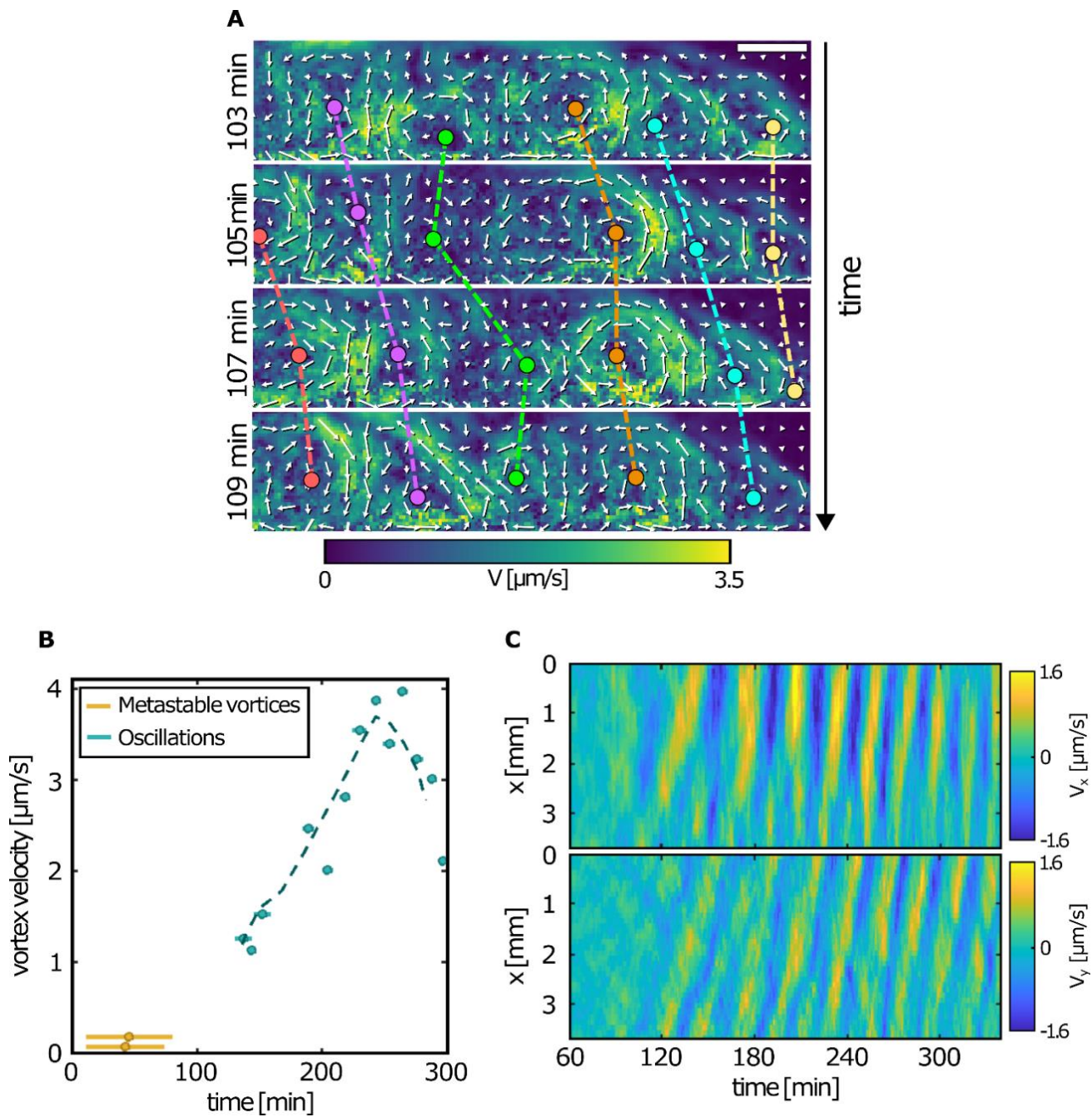

**Figure S11:** (A) Heatmap of the velocity magnitude versus time, highlighting large-scale vortex dynamics. Vortex centers (colored circles) and their trajectories (dashed lines) are overlaid. $C_{DNA} = 70\ ng/\mu l$, $L_{DNA} \approx 40\mu m$. Scalebar, 1 $mm$. (B) Vortex-center velocity magnitude over time for the standing-vortex state (yellow) and the traveling vortex in the oscillation regime (green). Velocities were obtained from linear fits of the vortex center position versus time. Error bars indicate the tracking intervals of vortices. (C) Space-time plots of the velocity $v_x$ and $v_y$ in the traveling vortices regime.

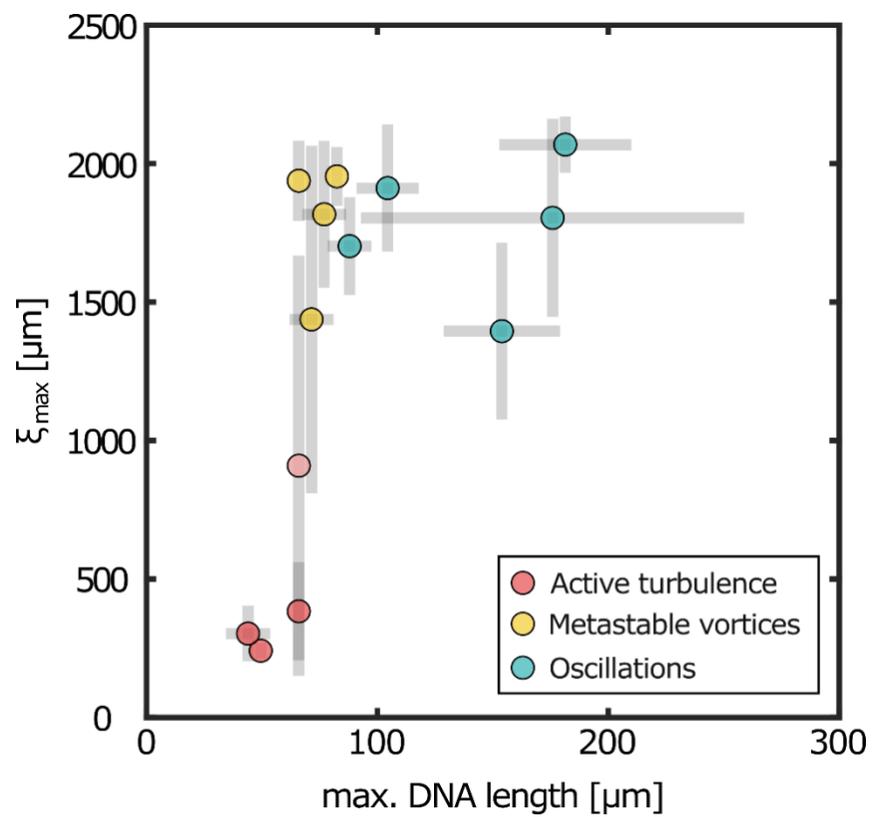

**Figure S12:** Maximal velocity-velocity correlation length as a function of maximal polymer length in the DNA distribution. Complementary to Fig. 3C.

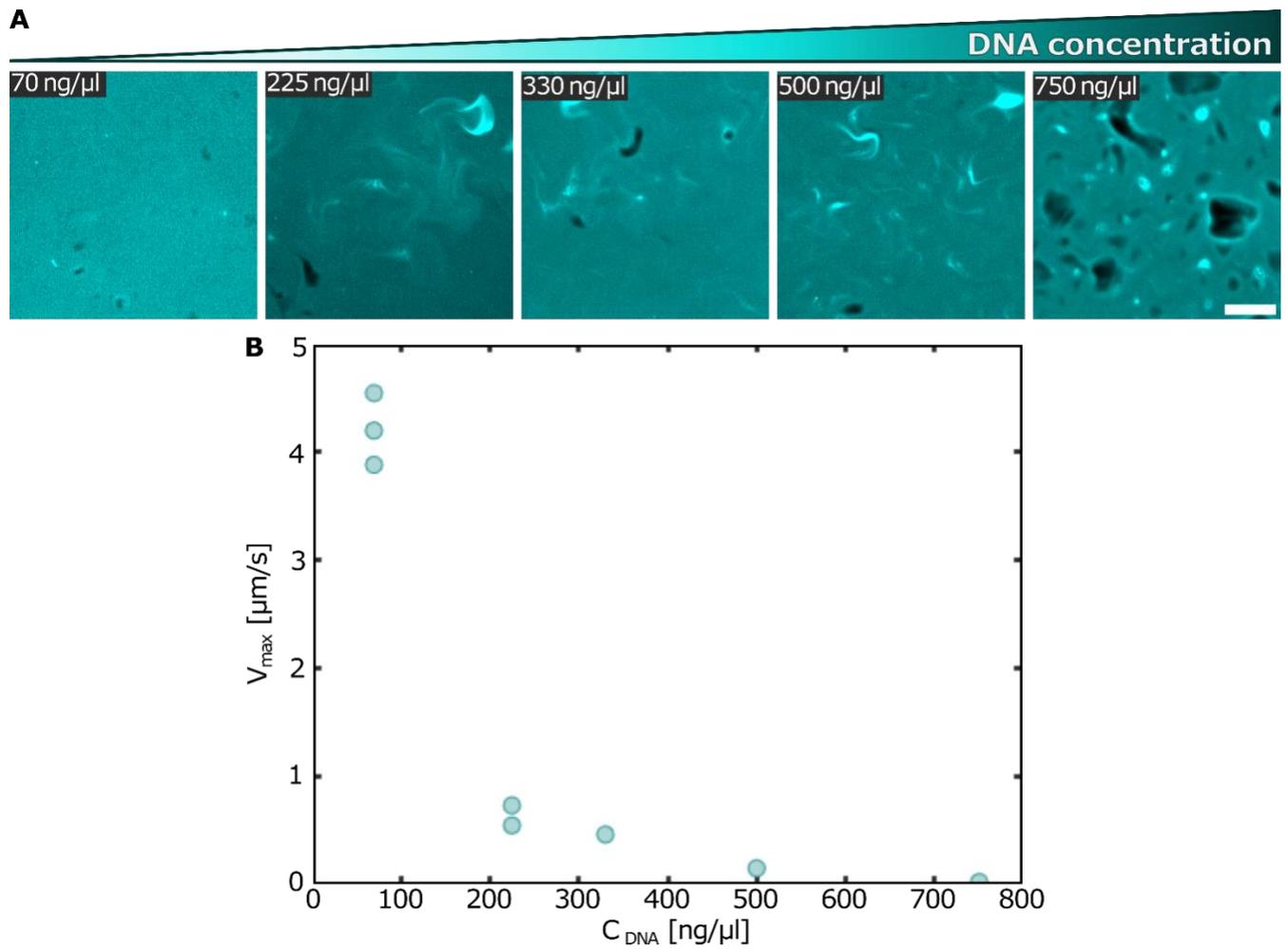

**Figure S13:** (A) Fluorescence images of λ-DNA in the active fluid, $L_{DNA} = 15\ \mu m$, at varying DNA concentrations. Dark domains indicate regions of reduced dye intercalation at high local DNA density. Scale bar, $250\ \mu m$. (B) Maximal velocity magnitude as a function of λ DNA concentration.

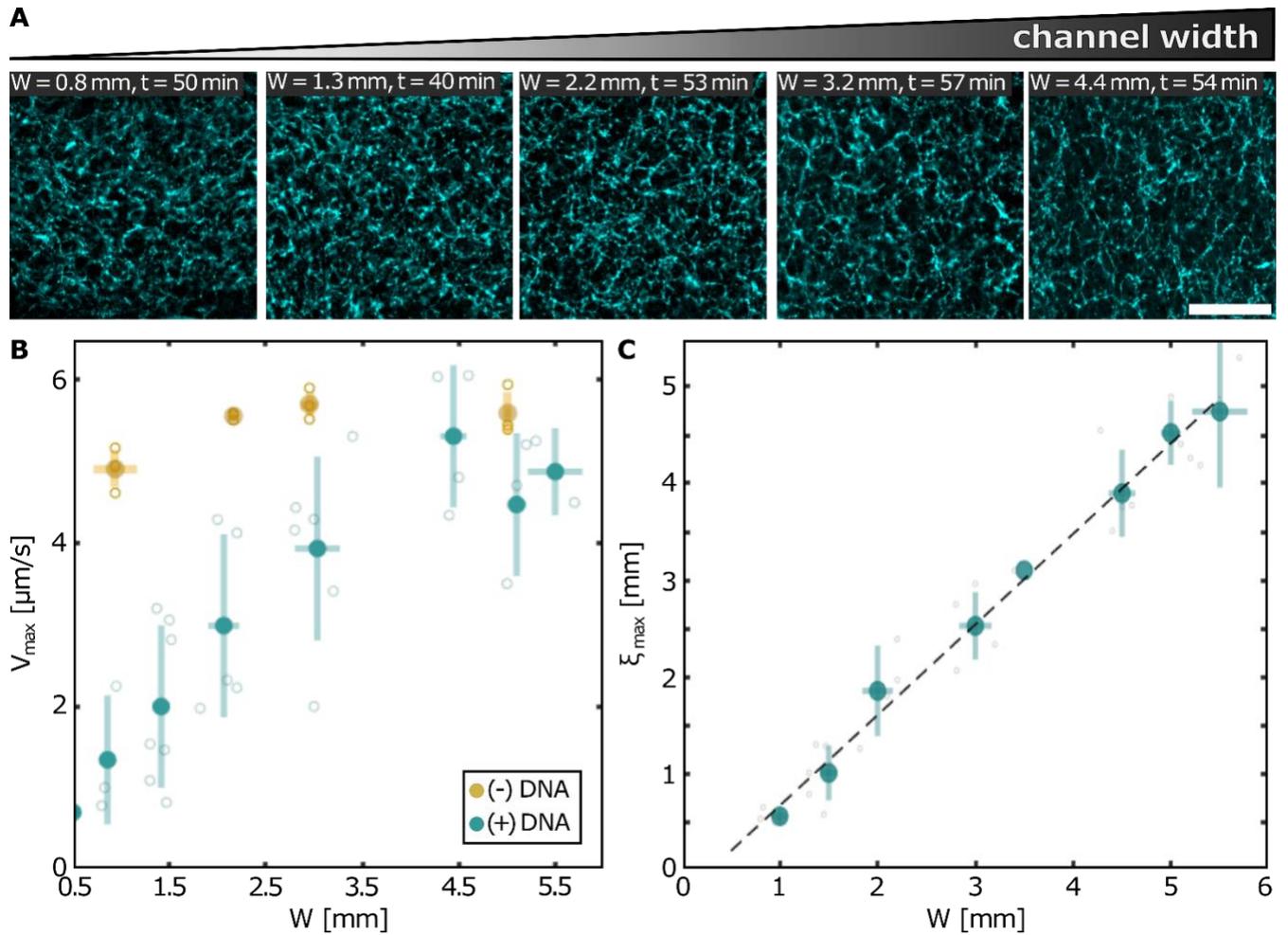

**Figure S14:** (A) Fluorescent images of DNA network morphologies at varying channel widths, $W$ at DNA concentration $C_{DNA} = 70 \, ng/\mu l$, and $L_{DNA} = 40 \mu m$. Scale bar, 250 $\mu m$. (B) Maximum velocity magnitude as a function of channel width, $W$, with (+) and without (-) DNA. Individual experiment scatter shown as empty markers. (C) Velocity-velocity correlation length as a function of channel width. Error bars represent standard deviation of channel width and correlation length measurements over three repeats. Individual experiments shown as gray empty markers.

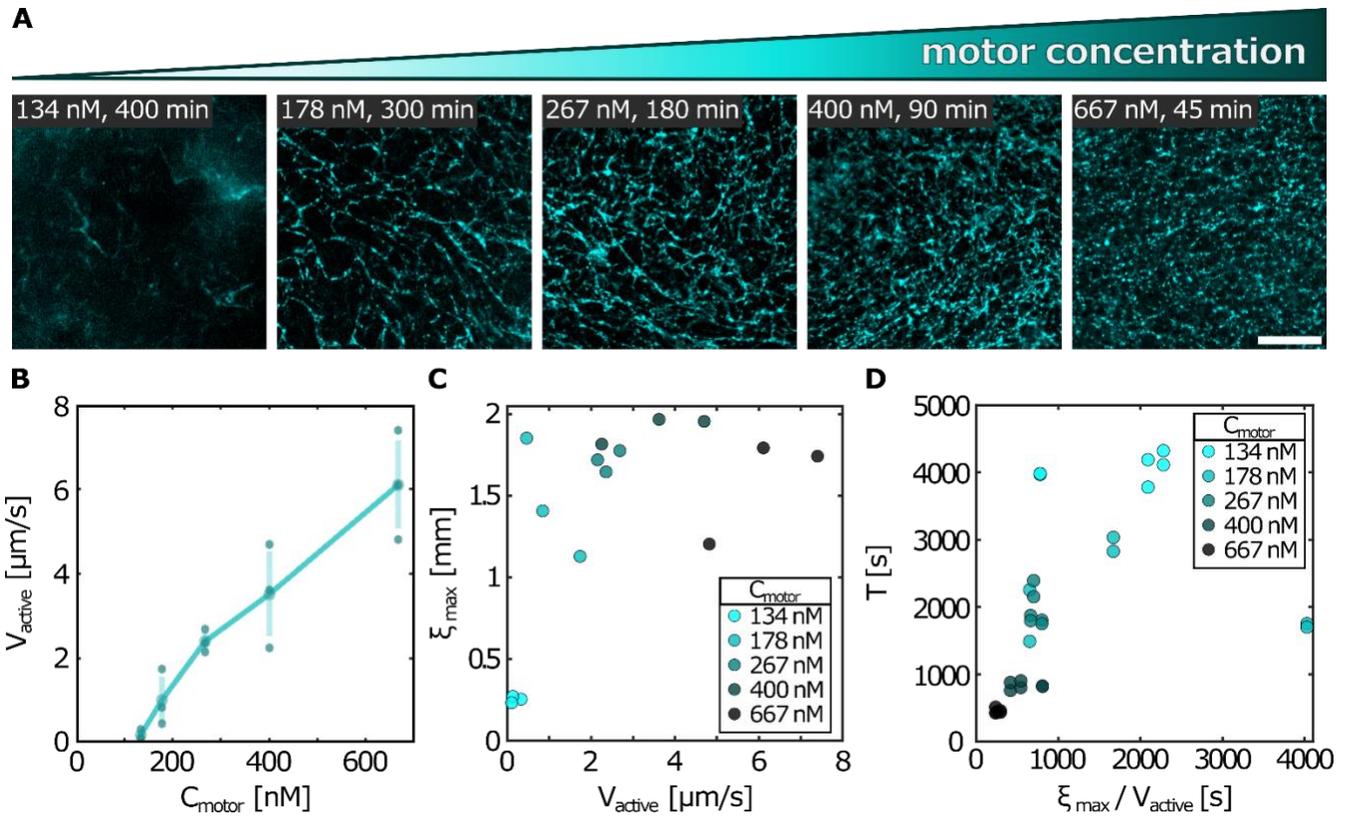

**Figure S15:** (A) Fluorescence images of the DNA network at varying motor concentrations. DNA concertation is $C_{DNA} = 70 \; ng/\mu L$ and DNA contour length is $L_{DNA} \approx 40 \mu m$. Images chosen to capture a formed network. Scale bar, $250 \mu m$. (B) Maximal local velocity magnitude as a function of motor concentration. Filled markers denote individual experiments, and vertical error bars show the standard deviation over three repeats per condition. (C) Maximal velocity-velocity correlation length $\xi_{max}$ as a function of maximal local velocity. Markers color-coded by motor concentration. (D) Oscillation period $T$ as a function of advection timescale.

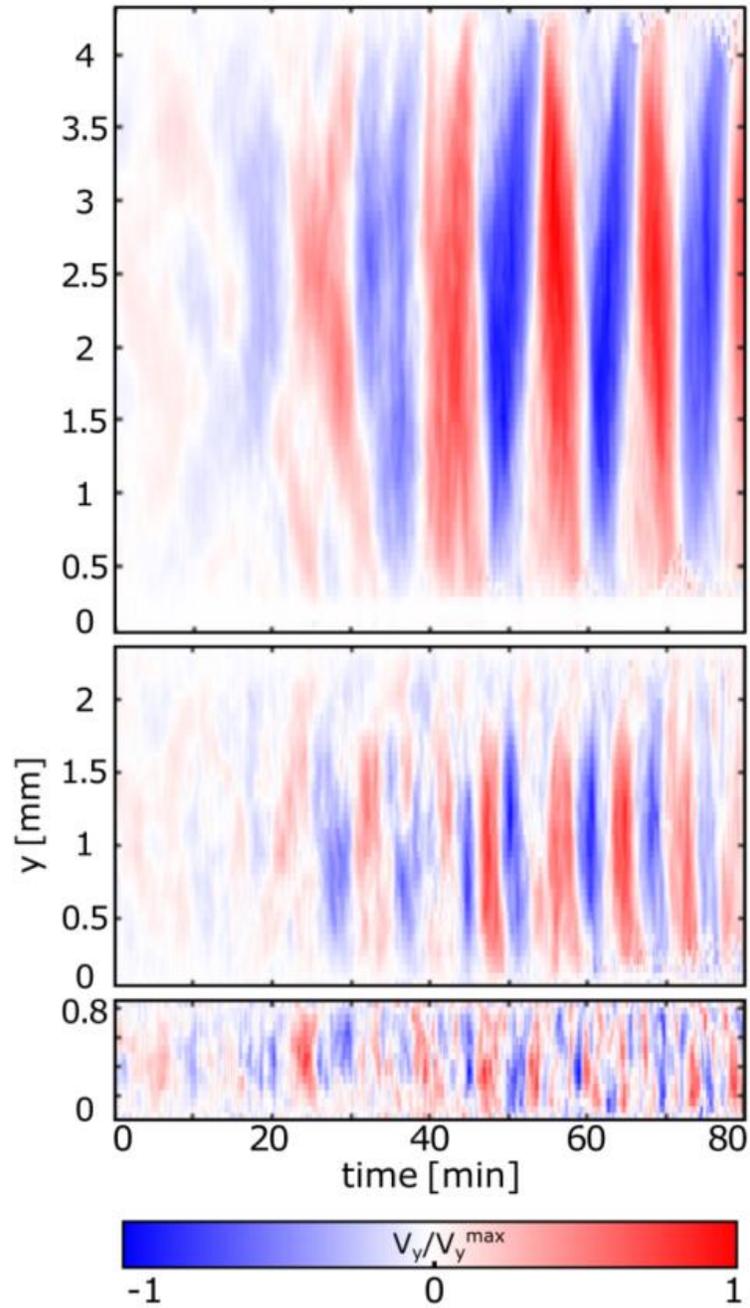

**Figure S16:** Space-time plots of the transverse velocity component $v_y$ along the short axis of the channel, $y$, sampled along the channel short axis $y$ different channel widths. Complementary to Fig. 4B.

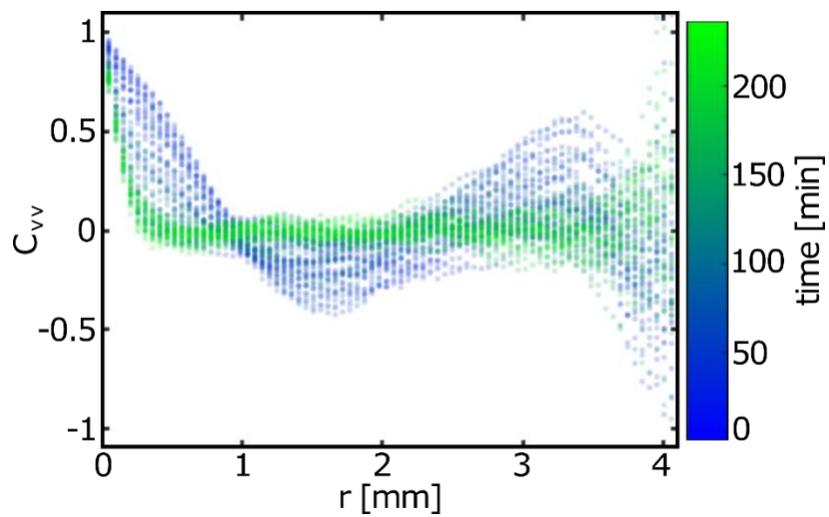

**Figure S17:** Velocity-velocity correlation function as a function of distance $r$, at different times. DNA concentration $C_{DNA} = 70\ ng/\mu l$, average DNA contour length $L_{DNA} \approx 40\ \mu m$.

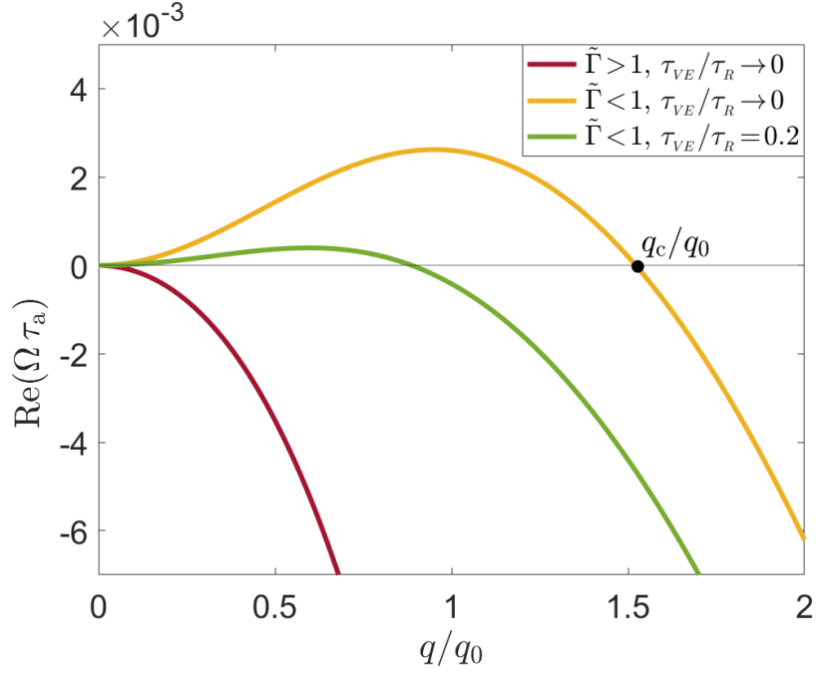

**Figure S18: The stationary (non-oscillatory) instability.** The real part of the stability spectrum $\Omega(q)$ of Eq. (10), $Re(\Omega\tau_a)$ (multiplied by $\tau_a = \Gamma l_a^2/\alpha$), is plotted vs. $q/q_0$ (with $q_0 \equiv \pi/W$, where $W$ is the system size) for vanishing and relatively small $\tau_{VE}/\tau_R$ values (and setting $\phi = 0$). For the entire range of $q$'s, the imaginary part of $\Omega(q)$ vanishes, $Im[\Omega(q)] = 0$, and is not plotted. The bottom (red) curve corresponds to $\frac{\tau_{VE}}{\tau_R} = 10^{-7}$ (i.e. vanishing viscoelasticity) and $\tilde{\Gamma} = 2 > 1$. The latter does not satisfy the inequality in (11) and indeed the system is stable for all $q$'s in this case, $Re[\Omega(q)] < 0$. The upper (yellow) curve corresponds to the same parameters as of the red curve, except that $\tilde{\Gamma} = 0.4 < 1$, this time satisfying the inequality in (11). As predicted, in this case a continuous range of unstable modes, corresponding to $0 < q < q_c$, emerge (featuring $Re[\Omega(q)] > 0$, $q_c/q_0$ is marked by the filled circle). The middle (green) curve corresponds to the same parameters as of the yellow curve, except that $\tau_{VE}/\tau_R = 0.2$ is finite, i.e., a viscoelastic response exists. As predicted, viscoelasticity tends to stabilize the system, leading to a shrinking instability range and hence to a growing length scale. The value of $\tau_{VE}/\tau_R$ has been selected such that $q_c \simeq q_0$, implying that the only unstable mode corresponds to $q_0$ (i.e., to the system size, recall that the wavenumber accessible to the finite systems satisfy $q > q_c$.) The parameters used in all curves are $q_0 l_a = \pi/20$, $l_h/l_a = 5$, and $\tau_R/\tau_a = 10$.

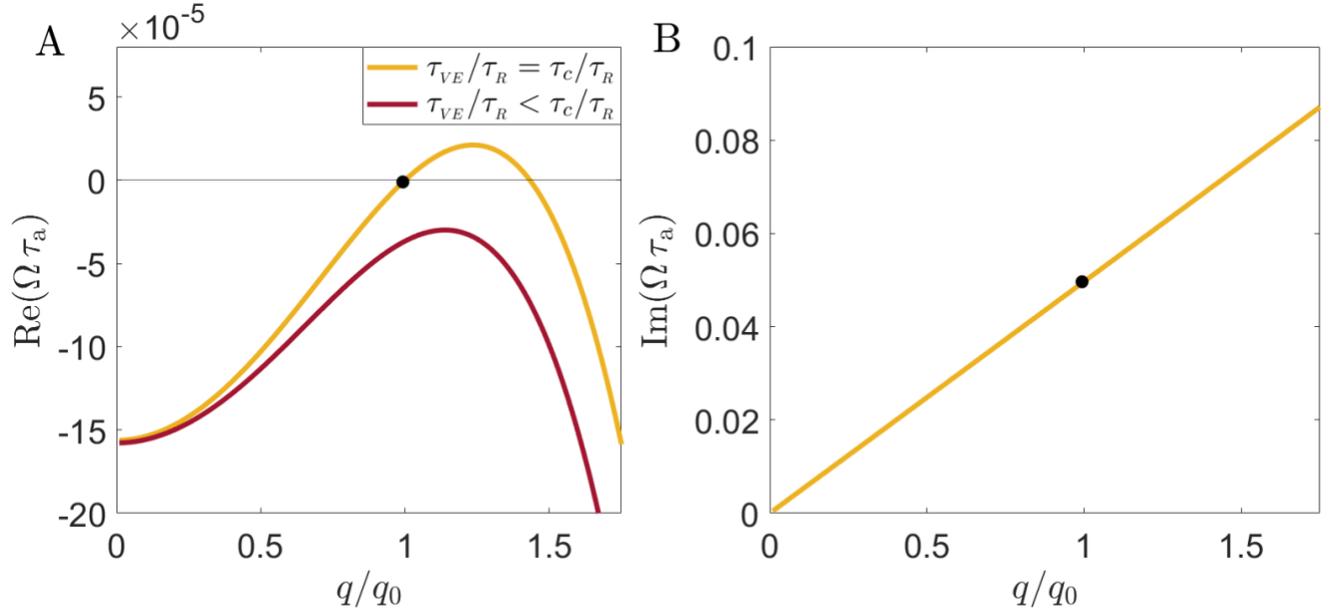

**Figure S19: The oscillatory (Hopf) instability.** (A) $Re(\Omega\tau_a)$ vs. $q/q_0$ as in Fig. S18, except that we use larger $\tau_{VE}/\tau_R$ values and set a larger $l_h/l_a = 30$, according to the expected length scale separation $l_a \ll W \ll l_h$, see text. The red curve is slightly below the onset of instability, corresponding to $\tau_{VE} = 0.99\tau_c$ where $\tau_{VE} = \tau_c$ corresponds to the value for which $q_0$ becomes unstable (see sup text). The latter is marked by a filled circle on the yellow curve plotted for $\tau_{VE} = \tau_c$. The onset of instability occurs between the two curves. (B) $Im(\Omega\tau_0)$ vs $q/q_0$ for $\tau_{VE} = \tau_c$, demonstrating the oscillatory nature of the instability. The filled circle marks the frequency of oscillations at the system scale, i.e., at $q = q_0$, see text.